\documentclass[a4paper]{article}
\usepackage{graphicx} % Required for inserting images
\usepackage{placeins}
\usepackage{hyperref}
\usepackage{booktabs}
\usepackage{parskip}
\usepackage{adjustbox}
\usepackage{authblk} % Authors and Affiliations

\usepackage{multicol} % For tables
\usepackage{multirow} % For tables
\usepackage{array} % For tables
\usepackage{subcaption} % for subfigures
\usepackage[margin=1.5cm]{geometry} % for margins
\usepackage{cancel} % for \cancel command
\usepackage{booktabs} % for \bottomrule and the like
\usepackage{amsmath}
\usepackage{amssymb}
\usepackage{tabularx}
\usepackage[numbers]{natbib}
\title{DecNefSimulator: A Modular, Interpretable Framework for Decoded Neurofeedback Simulation Using Generative Models\footnote{All DecNefSimulator code is publicly available at \url{https://github.com/AlexOlza/DecNefSimulator}}}

\author[1]{Alexander Olza \footnote{{Corresponding Author; alexander.olza@ehu.eus}\footnote{ORCID = 0000-0002-5987-6520}}}

\affil[1]{Intelligent Systems Group, University of the Basque Country (UPV/EHU); Donostia - San Sebastián, Spain}

\author[1]{Roberto Santana}
\author[2, 3]{David Soto}

\affil[2]{Consciousness Group, Basque Center on Cognition, Brain and Language (BCBL); Donostia - San Sebastián, Spain}

\affil[3]{Ikerbasque, Basque Foundation for Science; Bilbao, Spain}

\begin{document}
\maketitle

% Main text

% Numbered list
% Use the style of numbering in square brackets.
% If nothing is used, default style will be taken.
%\begin{enumerate}[a)]
%\item 
%\item 
%\item 
%\end{enumerate}  

% Unnumbered list
%\begin{itemize}
%\item 
%\item 
%\item 
%\end{itemize}  

% Description list
%\begin{description}
%\item[]
%\item[] 
%\item[] 
%\end{description}  

%\clearpage %%Remove this from your manuscript

\begin{abstract}
Decoded Neurofeedback (DecNef) is a promising non-invasive approach to brain modulation with wide-ranging applications in neuromedicine and cognitive neuroscience. However, progress in DecNef research remains constrained by subject-dependent learning variability, reliance on indirect measures to quantify progress, and the high cost and time demands of experimentation.

We present DecNefSimulator, a modular and interpretable simulation framework that formalizes DecNef as a machine learning problem. Beyond providing a virtual laboratory, DecNefSimulator enables researchers to model, analyze and understand neurofeedback dynamics. Using latent variable generative models as simulated participants, DecNefSimulator allows direct observation of internal cognitive states and systematic evaluation of how different protocol designs and subject characteristics influence learning.

We demonstrate how this approach can (i) reproduce empirical phenomena of DecNef learning, (ii) identify conditions under which DecNef feedback fails to induce learning, and (iii) guide the design of more robust and reliable DecNef protocols \textit{in silico} before human implementation.

In summary, DecNefSimulator bridges computational modeling and cognitive neuroscience, offering a principled foundation for methodological innovation, robust protocol design, and ultimately, a deeper understanding of DecNef-based brain modulation.
\end{abstract}

\section{Introduction} \label{sec:introduction}

Neurofeedback is a neuroscientific technique that enables participants to gain control over their brain activity by using real-time feedback, measured through neuroimaging, as a training signal~\cite{sitaram2017closed}. Decoded neurofeedback (DecNef) is a specialized form of neurofeedback that employs real-time reinforcement to induce a target brain state, even without explicitly requiring participants to think about specific contents or follow specific strategies to modulate their neural activity~\cite{shibata2011perceptual}. Unlike traditional neurofeedback, which relies on explicit self-regulation, DecNef leverages machine learning (ML)–based decoding and external rewards to guide neural adaptation. This approach allows for precise modulation of neural patterns while minimizing cognitive strategy biases~\cite{shibata2021mechanisms}.

The DecNef procedure consists of two main stages. In the classifier construction phase, fMRI data are collected while participants perform tasks or perceive stimuli associated with the target brain state. ML models are trained to recognize this neural pattern. This process is known as ``brain decoding''. In the induction phase, the trained classifier is applied to real-time fMRI signals to generate feedback that aims to implicitly guide participants toward the desired state through external reinforcement. This process is thought to promote neuroplasticity by modifying activity within specific brain regions or influencing broader network dynamics~\cite{shibata2019toward}. However, despite its promise, DecNef is currently subject to multiple challenges~\cite{shibata2021mechanisms}, as well as technical and methodological limitations, all of which constrain robustness and generalizability.

From an ML perspective, a domain shift between the classifier construction and induction stages is a fundamental issue~\cite{yamashita2008sparse,cortese2016multivoxel}. While the target pattern is typically identified in response to explicit stimuli, the corresponding cognitive state may produce a different neural signature during induction, when no stimuli are present~\cite{olza2025domain}. This mismatch can prevent participants from consistently achieving the intended brain state, reducing the efficacy of DecNef. In addition to these challenges related to brain decoding, behavioral and learning factors can further undermine training reliability.

Another critical concern involves the potential encouragement of maladaptive learning strategies: ways of maximizing feedback without achieving the target state. Since DecNef depends on the classifier’s ability to recognize patterns from its training set, brain states outside this range may yield unreliable or misleading outputs. This can be accentuated by well known limitations and vulnerabilities of ML models~\cite{moosavi2017universal,vadillo2022analysis}. Participants (who are unaware of the DecNef target) may exploit noise or spurious classifier responses to modulate their brain activity in unintended ways that hold no actual relationship with the target neural pattern, but also yield high rewards. In such cases, training would seem successful in the short term (as judged by the obtained DecNef reward), but would not reliably enhance the desired neural, cognitive or behavioural outcomes~\cite{stevens2021neural,heunis2020quality}. These practical issues connect to a deeper conceptual problem regarding what decoded signals truly reveal about brain function and representation.

Beneath these difficulties lies a core methodological question: what does ``successful decoding'' actually tell us about the brain? The rise of brain decoding popularized an inductive inference principle (or ``bridge principle'') from decodability to representation, known as the \textit{decoder's dictum}~\cite{ritchie2019decoding}: The assumption that, if we can reliably decode a variable from brain activity using a classifier, then that variable is explicitly represented in, and used by, the brain's own computations. Although widespread, this inference is contested, as it conflates what an external observer can interpret (neuroimaging data) with the information the brain itself functionally uses. Decoding success merely indicates that some information is available to the experimenter~\cite{ritchie2019decoding}, not that the decoded pattern constitutes causal information (a signal that genuinely influences downstream neural processes)~\cite{deWit2016neuroimaging,kragel2018representation}. Fundamentally, the decoder's dictum overlooks the inherently indirect nature of all neuroimaging measurements~\cite{deWit2016neuroimaging,logothetis2008what}. This collective challenge exposes a crucial empirical gap, emphasizing the need for approaches that move beyond correlation to capture the causal dynamics of information transfer within the brain~\cite{deWit2016neuroimaging}. Because DecNef uses decoder output as a training signal, any mismatch between decodability and causal representation directly affects what is reinforced. Thus, these conceptual limitations must be considered at every stage of DecNef conceptualization, protocol design, evaluation, and application. 

Beyond methodological and conceptual issues, a substantial proportion of participants simply fail to benefit from DecNef, often labeled as non-responders~\cite{alkoby2018can,weber2020predictors}. We hypothesize that these learning difficulties could be mitigated by addressing the foundational challenges outlined above. In this work, we investigate three candidate factors of apparent non-response: (i) the choice of alternative class used for the DecNef classifier, (ii) the participant’s initial cognitive state, and (iii) variability in the neural outcomes of regulation attempts.

Although the validity concerns mentioned above have already been discussed in the literature~\cite{thibault2017psychology,micolaud2018framework,ros2020consensus}, there is still no systematic framework that thoroughly elucidates the methodological limitations of DecNef. Furthermore, the real-time nature of DecNef complicates efforts to test new protocols in a principled and reproducible manner.

Here, we introduce a simulation framework that models DecNef training by replacing the human participant with an artificial agent. Using this framework, we analyze DecNef pitfalls from an ML standpoint and provide a controlled environment for exploring innovative DecNef protocols prior to real-life experimentation. The proposed framework is intended for \textit{in silico} evaluation and optimization of neurofeedback protocols prior to experimental validation in human studies, rather than for real-time closed-loop deployment. Accordingly, real-time computational constraints (e.g., inference latency or online deployment requirements) are outside the scope of this work, which focuses on pre-experimental protocol design and comparison. Evaluation is conducted in controlled simulation settings, including image data and synthetic fMRI, which provide a reproducible testing environment. In this context, our contributions are threefold: (i) we introduce DecNefSimulator, a modular simulation framework that replaces human participants with latent variable generative models; (ii) we use it to analyze how alternative-class choice, initial conditions, and stochastic regulation shape DecNef learning and apparent non-response; and (iii) we demonstrate how this framework can diagnose maladaptive learning and guide protocol design \textit{in silico} before human experiments. We believe that continued methodological development, driven by advances in ML, could help overcome current barriers to applicability, unlocking new opportunities for non-invasive brain modulation. Strengthening DecNef's methodological foundations is therefore essential to ensure its robustness, interpretability, and reproducibility, and we propose our simulation framework as a step toward that goal.

\section{Related work}

\subsection{Current approaches to DecNef: theory and practice}

Current approaches to DecNef are typically organized into two main stages. In the classifier construction phase, neuroimaging data are acquired while participants perform tasks or are exposed to stimuli designed to evoke the target brain state, as well as other control conditions, and ML methods are trained to discriminate the brain activity patterns associated to the target state from the alternative conditions. Typically, binary supervised classifiers are employed~\cite{taschereau2020conducting}. Some works mention the possibility of using multiclass classifiers~\cite{cortese2017decoded,koizumi2016fear,shibata2019toward}, although the output is customarily binarized for feedback computation~\cite{koizumi2016fear}. Either way, it is essential to remember that alternative categories hold equal importance as the target category, as they impact the classifier's specificity~\cite{taschereau2020conducting}. Linear models, such as sparse logistic regression or linear Support Vector Machines (SVMs), are the most commonly adopted~\cite{yamashita2008sparse, shibata2011perceptual}. Although non-linear classifiers have been considered in principle, real-time implementation has been limited due to speed and interpretability concerns~\cite{shibata2019toward, taschereau2020conducting}. Neural networks are rarely considered due to the same reasons.

In the subsequent induction phase, the classifier is integrated into an online fMRI feedback loop. Brain scans are acquired and decoded in real-time, and the probability of belonging to the target class is scaled into a reward function, often implemented via a visual cue (such as adjustments in the size of a disc or bar)~\cite{dealmeida2011vsimg}. This transformation is usually linear, though adaptive reward scaling is under exploration in some studies~\cite{cortese2017decoded, shibata2019toward}.

Crucially, participants are not provided with any information about the purpose of  DecNef. Instead, the paradigm relies on implicit reinforcement, and participants must explore and exploit the mental strategies that are more efficient at maximizing reward feedback. Thus, DecNef seeks to encourage the modulation of brain activity toward a pattern that maximizes the probability of the target according to the classifier, without a priori knowledge of such target. Across repeated sessions, such closed-loop reinforcement is hypothesized to induce neuroplastic changes both within specific cortical regions and at the level of large-scale network interactions~\cite{shibata2019toward,shibata2021mechanisms}.

Some recent developments highlight the potential of adaptive discriminators that adjust to non-stationary brain signals~\cite{abdennour2026enhancing, klawiter2026sequential}. Additionally, causal connectivity is recently being explored as a way to test and explain neurofeedback success~\cite{arab2024whole}. Nevertheless,  these are not yet widely deployed in DecNef studies, which still favor computationally efficient methods due to the real-time nature of the experiments.

Theoretical models of DecNef mechanisms from the point of view of ML are scarce, but in growing development. As explored in previous research~\cite{lawrence2014selfregulation, lubianiker2022neurofeedback} DecNef has been framed as a Reinforcement Learning (RL) task.  In~\cite{lubianiker2022neurofeedback}, the authors view DecNef as a RL scenario with agent-environment conflation, where the same brain that selects the actions (thus, the agent) is also the brain whose states will be changed by that action (thus, it is also the environment). Due to the implicit nature of DecNef, participants may explore a wide range of  strategies to maximise reward. Hence, actions in DecNef are complex and multifactorial (i.e. imagining a person dancing in a concert)  but the feedback is tied to the aggregated activity,  leaving the trainee to figure out which features (music, movement...) are responsible for the successful feedback, a phenomenon known as \textit{credit assignment problem} in RL jargon. A number of works also attempt to identify underlying representation learning methods in the human brain, connected to solving the curse of dimensionality in RL-like tasks~\cite{niv2015reinforcement}.

Another relevant area of research is trying to identify predictor variables of neurofeedback success~\cite{haugg2021predictors,alkoby2018can}.
However, as those articles point out, there is no agreement on how to quantify success (i.e. based on feedback improvement, behavioural outcomes, or neurophysiological examination). The heterogeneity in the presentation of results in neurofeedback literature complicates the analysis, although reporting standards are being encouraged~\cite{ros2020consensus, tursic2020systematic}.

\subsection{Simulations of human learning and human brain functioning}

\def\arraystretch{1.5}
\begin{table}[t]
\centering
\renewcommand{\tabcolsep}{6pt}
\resizebox{\textwidth}{!}{
\begin{tabularx}{\textwidth}{
>{\raggedright\arraybackslash}p{2.6cm}||
>{\raggedright\arraybackslash}X|
>{\raggedright\arraybackslash}X|
>{\raggedright\arraybackslash}X|
>{\raggedright\arraybackslash}X}
\toprule
\textbf{Aspect} & \textbf{Oblak et al. (2017)} & \textbf{Shibata et al. (2019)} & \textbf{Annicchiarico et al. (2025)} & \textbf{DecNefSimulator (Ours)} \\
\midrule\midrule
\textbf{Task} & Orientation & Orientation & Toy & Semantic category induction \\
\midrule
\textbf{Brain model} & Cube of voxels (visual cortex model) & ``Biologically plausible'' neural network & Array of 5 integer states & Latent variable generative model \\
\midrule
\textbf{State space} & Set of 8 orientations (Gabor patches) & Set of 8 orientations (Gabor patches) & Set 5 of integers & Generator manifold (with semantic structure) \\
\midrule
\textbf{Subject characteristics} & Physiological filters & Physiological filters & Prior beliefs, confidence in feedback & Prior beliefs, confidence in feedback, decision-making preferences \\
\midrule
\textbf{Neuroplasticity} & No & Yes & No & No \\
\midrule
\textbf{Cognition} & Yes & No & Yes & Yes \\
\midrule
\textbf{Reward} & Target-class likelihood & Binary & Discretized normal around target state & Target-class likelihood \\
\midrule
\textbf{State transitions} & Multiplicative effect of feedback changes on each voxel  & Hebbian rule modification of synaptic weights & Probabilistic transition to adjacent states using active inference & Continuous latent-space traversal through parametrized learning strategies \\
\midrule
\textbf{Main goal} & Test effect of feedback presentation scheduling and physiological filters & Test the neuroplasticity hypothesis as the main mechanism in DecNef & Analyze impact of subject's beliefs and feedback quality & Provide an end-to-end framework to develop new DecNef protocols \textit{in silico} \\

\bottomrule
\end{tabularx}
}
\vspace{0.5em}
\caption{Comparison of artificial DecNef simulations. 
}
\label{tab:comparison-RW}
\end{table}

The use of computational tools to model human learning in neurofeedback is emerging as a promising strategy to test paradigms before implementation with human participants~\cite{cui2024opinion, davelaar2022multistage}. Yet, to our knowledge, there are very few works implementing actual neurofeedback simulations beyond mere theoretical considerations.

The few existing DecNef simulations, in which the human participant is replaced by an artificial model, have pursued different goals: replicating experimental results from prior DecNef empirical research~\cite{oblak2017selfregulation, shibata2019toward}, testing the neuroplasticity of biologically inspired neural networks~\cite{shibata2019toward}, and assessing the impact of subject characteristics on neurofeedback performance, either through physiological filters~\cite{oblak2017selfregulation, shibata2019toward} or through prior beliefs and habits~\cite{annicchiarico2025activeinferenceperspectiveneurofeedback}.

A pivotal precursor to these simulations was the DecNef experiment conducted in 2011 by Shibata et al.~\cite{shibata2011perceptual}, which showed that participants could learn to evoke brain activity associated with the orientation of Gabor patches, and that DecNef training triggered subsequent perceptual learning for the evoked Gabor orientation. This study provided the first empirical evidence that DecNef could shape brain activity by reinforcing neural patterns linked to simple visual features. Building on this foundation, subsequent studies~\cite{oblak2017selfregulation, shibata2019toward} developed DecNef simulations which replicated the behavioral outcomes of Shibata et al.

Work presented in~\cite{oblak2017selfregulation} constitutes one of the first DecNef simulations, designed around the orientation task established by~\cite{shibata2011perceptual}. The brain was modeled as a parametrically defined cube of voxels representing the early visual cortex (V1), in which a subset of voxels were designed to display orientation-specific activity. The state space was continuous, defined by real-valued voxel intensities. Subject brain properties were incorporated through physiological filters such as the hemodynamic response function. Neuroplasticity was not explicitly modeled, but cognitive processes were indirectly represented by linking voxel responses to orientation features. Feedback was computed as the likelihood of the target orientation class, and voxel-wise reinforcement learning was used as the update rule, modeling voxel-wise regulation as a multiplicative effect proportional to feedback change. The main goal was to provide a proof of concept that DecNef induction could be simulated in silico under controlled conditions.

A follow-up study~\cite{shibata2019toward} also simulated the orientation task, but with a biologically inspired neural network. Here, the brain model was based on a self organizing map~\cite{kohonen2013essentials} of orientation-selective neurons, resulting in a discrete set of stable orientation states. Physiological filters were again used to approximate subject variability. In contrast to~\cite{oblak2017selfregulation}, neuroplasticity was explicitly modeled through synaptic weight changes, updated via a Hebbian learning rule modulated by binary reward signals. Cognition was not explicitly represented, and reward was simplified to a binary feedback signal indicating whether the converged pattern matched the target orientation. The goal of this study was to mimic neuronal plasticity from a mechanistic perspective, highlighting how DecNef could strengthen orientation-selective synapses.

More recently, authors of~\cite{annicchiarico2025activeinferenceperspectiveneurofeedback} approached DecNef simulation from the perspective of active inference in a ``toy task'' with one cognitive dimension. The brain model was defined in terms of discrete states rather than voxel activations or neural populations, and the state space was finite and integer-valued, lacking semantic continuity. Subject's prior beliefs and habits were incorporated using a discrete Partially Observable Markov Decision Process (POMDP)~\cite{spaan2012pomdp, maisto2025active}. 
Rewards were implemented as discretized normal distributions around the target state, and active inference was used as the learning rule. The main goal was to disentangle the contributions of cognitive strategies and prior beliefs to DecNef performance. For instance, the authors analyzed the relationship between feedback noise, subject's trust/mistrust around DecNef feedback and learning outcomes.

Taken together, these studies demonstrate the potential of DecNef simulations to investigate different aspects of learning and brain functioning. Yet, their applicability remains constrained by crucial limitations. Both~\cite{oblak2017selfregulation} and~\cite{shibata2019toward} rely on highly specific brain models that cannot be generalized beyond orientation selectivity. In contrast, the framework in~\cite{annicchiarico2025activeinferenceperspectiveneurofeedback} is limited by its discrete and finite mental state space, as well as the absence of semantic structure linking different cognitive states. These shortcomings highlight the pressing need for adaptable and task-flexible simulation frameworks.

Another factor to consider is the similarity of the DecNef simulation's data space with the fMRI data on which the real-world application is based. In relation to this, a relevant research area is the generation of synthetic fMRI data. Specifically, the synthesis of fMRI from images (emulating brain activity evoked by visual stimuli) is an increasingly active field. Recent works such as MindSimulator~\cite{bao2025mindsimulator} and SynBrain~\cite{mai2025synbrain} take images as input and return fMRI, enabling the creation of large annotated databases, where the label is an image and the associated datum is an fMRI scan. Recently, a unifying predictive model for subject-averaged brain activity in reaction to images, sound and video was released~\cite{dascoli2026foundation}. This model can also be used to generate synthetic fMRI data for a generic (``average'') subject. However, integration between fMRI synthesis and DecNef simulation remains unexplored.

Unlike these models that are either task-specific or cognitively coarse, our work (outlined in Table~\ref{tab:comparison-RW}, and described thoroughly in the sections below) utilizes latent variable generative models --- ML models that learn a low-dimensional latent space through an encoder and generate data through a decoder ---, with the resulting latent spaces capturing the semantic structure of the data. This approach allows for more nuanced representations of subject's characteristics and enables modular simulation approaches, where the nature of the artificial participant is decoupled from the DecNef task to be simulated. Several works have studied parallelisms between latent variable generative models and the human brain~\cite{breedlove2020generative, perl2020generative,perl2023low-dimensional, ozcelik2023natural, kamitani2025visual}, paving the way for their use in DecNef simulation frameworks. Building on this foundation, we introduce a novel approach that leverages generative models to reproduce, understand and analyze DecNef processes from a ML perspective. Our DecNefSimulator framework supports a variety of data modalities, including images and synthetic fMRI data, as shown in the following sections.
\FloatBarrier

\section{General DecNefSimulator framework for DecNef simulation}
This section outlines the broad DecNefSimulator framework, listing the main components and introducing the necessary notation in a general setting, making parallels to real-life DecNef. Each component can be chosen and replaced independently in a modular and flexible manner. For a concrete example, refer to Section \ref{sec:model-definition}.

\subsection{Static components: Generator (a.k.a. artificial  ``participant'') and Classifier}
 
In DecNefSimulator, the human participant is replaced by an artificial latent variable generative model $\mathcal{G}$ (the ``generator'', a.k.a ``participant' or ``subject'') with an internal, hidden space of cognitive states (the latent space $Z\subset \mathbb{R}^{m}$)  which can be indirectly ``observed'' using a proxy representation in an observable data-space $X$ contained in $\mathbb{R}^{n}$.

The architecture of $\mathcal{G}$ consists of an encoder $E_{\mathcal{G}}: \mathbb{R}^n \rightarrow Z$ and a decoder $D_{\mathcal{G}}: Z \rightarrow \mathbb{R}^n$.  To learn the structure of the latent space $Z$, $\mathcal{G}$ is trained in an unsupervised setting using a data-set $\{X_\mathcal{G}\}$ from any data modality of domain (i.e. images, voxel intensities resembling fMRI, or any other). Training $\mathcal{G}$ establishes a relationship between latent cognitive states in $Z$ and their observable representation in $\mathbb{R}^{n}$.

As a generative model, $\mathcal{G}$ can produce novel observations in $\mathbb{R}^{n}$ not present in the training dataset $\{X_\mathcal{G}\}$ but consistent with its learned underlying distribution. The set of all points in $\mathbb{R}^{n}$ that can be generated by the decoder defines the data space $X = \{ D_{\mathcal{G}}(z) \mid z\in \mathbb{R}^m \}$. In some fields of ML, $X$ is called the generator manifold. In other words, the set of all plausible observations of the subject's cognitive states, $X$, constitutes the data space or generator manifold.

Identical to the real-life DecNef setup, a supervised probabilistic classifier $\mathcal{D}$  is trained on a labeled dataset of indirect observations of the subject's cognitive states when exposed to known and controlled stimuli, $\{X_\mathcal{D}, Y_\mathcal{D}\}$. In real-life DecNef, $x_\mathcal{D}$ are fMRI recordings. However, as mentioned above, DecNefSimulator admits working with other data modalities (as long and $X_\mathcal{D}$ and $X_\mathcal{G}$ belong to the same modality). The labels $y_\mathcal{D}$ take values from a limited set, with each label semantically identifying the stimulus category. Typically, the classifier is restricted to a binary setting with a target class $y^\star$ and an alternative class $y^{\mathrm{alt}}$.

The sets of internal cognitive states associated to the target and alternative classes are denoted $Z^\star$ and $Z^{\mathrm{alt}}$, respectively. The representation of those sets in the data space is denoted $X^\star$ and $X^{\mathrm{alt}}$.  $\{X_\mathcal{D}\}$ only contains samples from $X^\star$ and $X^{\mathrm{alt}}$, and is unaware of their latent representation in $Z$.

In the standard supervised learning setting, the classifier $\mathcal{D}$ estimates the conditional probability $P(y \mid x)$ by mapping observable representations of cognitive states to the semantic labels provided during training. However, $\{X_\mathcal{G}\}$ contains observations not associated with either of the conditions seen for classifier training, which implies that $Z$ and, in turn, $X$, extends out of the classifier's knowledge.

\subsection{Dynamic components: Modeling learning strategies and decision-making}

The iterative nature of the induction stage from real-life DecNef is expressed in DecNefSimulator simulations using temporal subscripts. For instance, latent cognitive state at time $t$ is denoted $z_t$, and its projection to $X$ (its observable representation) is $x_t = D_\mathcal{G}(z_t)$. The sequences $\{z_t\}_{t=0}^T$ and $\{x_t\}_{t=0}^T$ are referred to as the cognitive trajectory and observable trajectory, respectively.

At each time step, the classifier $\mathcal{D}$ defined above estimates the probability that $x_t$ belongs to the target class $y^\star$ (in opposition to the alternative-class $y^{\mathrm{alt}}$) as $p_{t+1} = P(y = y^\star \mid x_t)$. This probability constitutes the feedback or reward signal at time $t+1$.

To simulate cognition, learning strategies (a.k.a. update rules) are denoted as $\mathcal{L}(z_t, p_{t+1}, \ldots)$. The update rule(s) characterize the subject's beliefs and decision-making behavior, translating neuroscientific knowledge or hypotheses into modeling assumptions. At minimum, $\mathcal{L}$ depends on the current cognitive state $z_t$ and, under the main DecNef assumption, on the feedback signal $p_{t+1}$ elicited by $z_t$, but it may also incorporate additional factors such as memory of previous feedback values. At each time step, the participant applies $\mathcal{L}$ to update its internal state:  $z_{t+1} = \mathcal{L}(z_t, p_{t+1}, ...)$.

\subsection{Summary of the notation}

\begin{figure}[h]
    \centering
    \includegraphics[width=\linewidth]{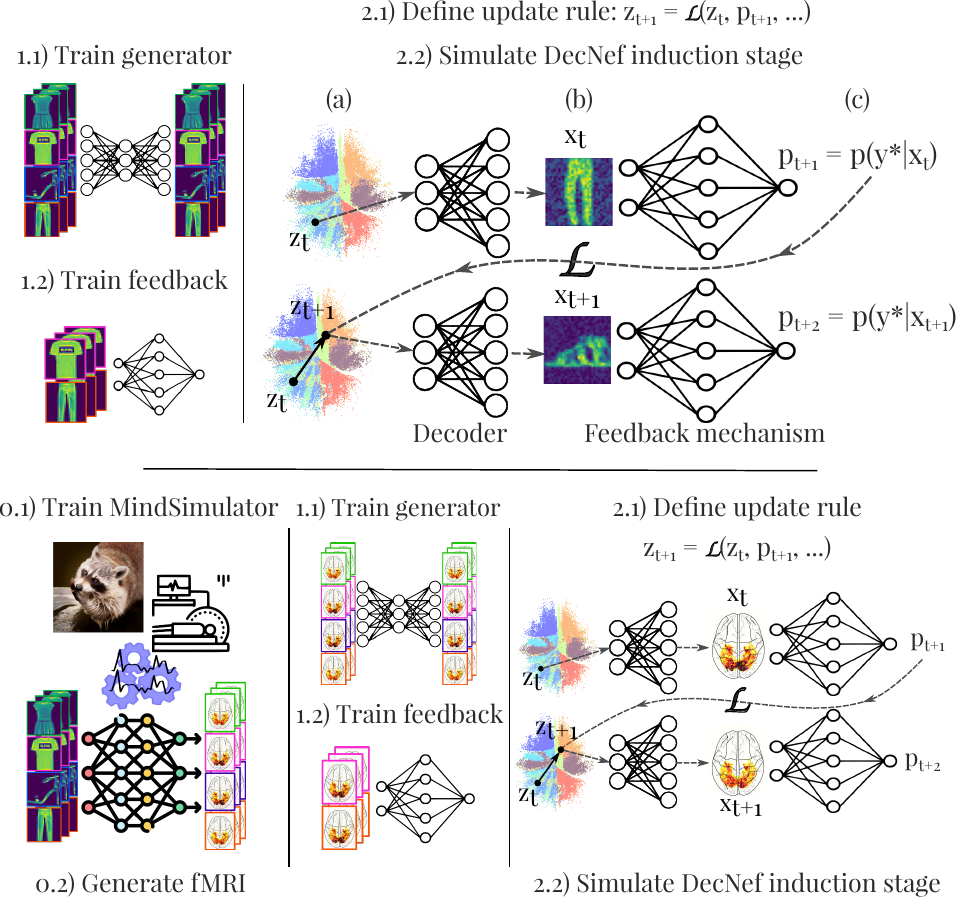}
    \caption{Two examples using the DecNefSimulator framework with different data modalities --- images (top) and synthetic fMRI (bottom) ---. Step one: Train the static framework components (generator and feedback mechanism). Step two: Use the static components and assigned learning dynamics (i.e. update rule $\mathcal{L}$) to simulate DecNef iterations. In step two, each DecNef iteration is composed of three actions: (a) sampling a point $z_t$ in the latent space, (b) projecting $z_t$ to the observable space using the decoder to obtain $x_t$, and (c) computing the next feedback $p_{t+1}$. After each iteration, the cognitive state $z$ is updated by moving through the latent space using $\mathcal{L}$. In case of synthetic fMRI, data generation (step zero) is independent from the rest of the pipeline.}
    \label{fig:generic-diagram}
\end{figure}

Table \ref{tab:notation} collects the notation introduced above for easier reference, and Figure \ref{fig:generic-diagram} depicts the basic pipeline of DecNef simulation.

As Figure \ref{fig:generic-diagram} shows, the simulation starts from a point $z_0$ in the latent space. The decoder $E_\mathcal{G}$ computes the corresponding $x_0$ in the data space, and the classifier computes the likelihood that $x_0$ belongs to the target-class, denoted $p_1$. The feedback signal based on $p_1$ informs the decision to modulate the internal state according to the subject's learning strategy $\mathcal{L}$, producing the new state $z_{1} = \mathcal{L}(z_0, p_1, ...)$. This dynamic continues in a closed loop until $T$ iterations are completed.

\def\arraystretch{1.6}
\begin{table}[h!]
\begin{tabular}{m{0.1\textwidth}|m{0.42\textwidth}|m{0.38\textwidth}}
    & \textbf{Real-life DecNef} & \textbf{Simulation} \\ 
    \hline \hline
    \multicolumn{3}{c}{Data modality  or domain, and static components of DecNef} \\
    \hline \hline
    $X \subset \mathbb{R}^{n}$ & fMRI (or other neuroimaging modality)& Any data modality (e.g. images, synthetic fMRI voxels, etc.)\\
    %\hline
     $x \in X $ & \multicolumn{2}{l}{A specific sample from the corresponding data modality}\\
    \hline
    $y^\star$ & \multicolumn{2}{l}{Semantic interpretation of the target class (label for desired cognitive states)} \\
    $y^{\mathrm{alt}}$ & \multicolumn{2}{l}{Semantic interpretation (or label) for the alternative class known to the classifier} \\
    \hline
    $X^\star \subset X$ &\multicolumn{2}{l}{Observable representation of the set of desired cognitive states}\\
    $X^{\mathrm{alt}} \subset X$ & \multicolumn{2}{l}{Observable representation of the set of alternative cognitive states} \\
    \hline
    $\mathcal{D}$ & \multicolumn{2}{l}{Supervised probabilistic classifier estimating $p(y = y^\star \mid x)$} \\
    $\{ x_\mathcal{D}, y_\mathcal{D} \}$ & \multicolumn{2}{l}{Labeled dataset for classifier construction}\\
    
    \hline\hline
    \multicolumn{3}{c}{Participant or subject} \\
    \hline\hline
    Subject & Human & Latent variable generative model $\mathcal{G}$ \\
    & & $Z \subset \mathbb{R}^{m}$: latent space \\
    & & $E_\mathcal{G}: X \rightarrow Z$: encoder \\
    & & $D_\mathcal{G}: Z \rightarrow X$: decoder \\    
    \hline
    Internal cognitive states & Not directly observable & $z \in Z$: a specific cognitive state \\
    & &  $Z^\star \subset Z$: The subset of cognitive states to be encouraged\\
    \hline 
    Internal dynamics & \multirow{3}*{Not observable} & \multirow{3}*{$\mathcal{L}$: Explicit model of the particular subject} \\
    Beliefs  & & \\
    Decision-making  & & \\
    \hline\hline
    \multicolumn{3}{c}{Sequence of real-time observations used for feedback (``observable trajectory'')} \\
    \hline
    $\{x_t\}_{t=0}^{T}$ & Using fMRI & Using $D_\mathcal{G}$\\
    \hline\hline
    \multicolumn{3}{c}{Sequence of internal cognitive states induced by DecNef in real time  (``cognitive trajectory'')} \\
    \hline
    $\{ z_t\}_{t=0}^{T}$ & Not observable & Observable \\
    \hline
\end{tabular}

\caption{General DecNefSimulator framework: Summary of the notation.}
\label{tab:notation}
\end{table}

\FloatBarrier
\section{Systematic simulation of DecNef experiments using DecNefSimulator}

 As a proof of concept, we present a collection of DecNef simulations conducted on a Computer Vision dataset using a specific model for the artificial participant. The complete experimental setting is directly applicable to other domains of data, but using images as the observable representation of the cognitive states offers complete transparency, facilitating demonstration. We then simulate the same task using synthetic fMRI data obtained from MindSimulator~\cite{bao2025mindsimulator}, increasing similarity to real life DecNef dynamics. This synthetic fMRI dataset models brain activity patterns elicited from visual perception of the same Computer Vision dataset used for our first round of experiments.
 
 The generative model and the update rule or learning strategy, which determine the characteristics of the artificial subject, are the backbone of DecNefSimulator simulation. As such, they must balance model simplicity and neuroscientific plausibility. Different theories of brain functioning and human learning behaviour will translate to different model characteristics, replacing the equations presented in Section \ref{sec:model-definition}.%By choosing a generator and an update rule, we recreate a traditional DecNef loop and we demonstrate its behaviour. \\%, highlighting and explaining the potential pitfalls of this procedure.

 The feedback presentation scheduling and computation method, which constitute the DecNef protocol, can also be changed freely. Here we used a binary classifier and a semi-continuous reward presentation averaging the feedback on three subsequent observations of the cognitive state, as customary in many DecNef studies. 

\subsection{Model definition}\label{sec:model-definition}

In this example, we used a Variational Autoencoder (VAE)~\cite{kingma2019introduction} as a generator, and we chose a learning strategy $\mathcal{L}$ based on two main assumptions: (1) that transitions in the latent space are inherently stochastic, reflecting variability and uncertainty in how the participants' mental strategies translate into neural states. (2) that the degree of exploration is modulated by feedback, such that increasing reward favors exploitation (reduced exploration), whereas decreasing reward promotes exploration. Since the DecNef participant is instructed to maximize the reward signal, the learning process requires short-term memory of recent feedback. Accordingly, the model assumes (3) that a significant drop in the reward signal leads the participant to revert to the previous state and attempt regulation again from that state.

\subsubsection{Generative model}
For the generator, we use a VAE. The latent space of the VAE contains the subject's possible cognitive states in the absence of external stimuli (i.e. during mental imagery). The generator manifold $X = \{ D_\mathcal{G}(z) \mid z \in \mathbb{R}^{m} \}$ defines the data space, and contains the observable representation of those cognitive states (which, in empirical DecNef, would be fMRI scans --- here, we use either visual images or synthetic fMRI scans).

Thus, $z$ are points in the latent space of the VAE (internal cognitive states), and $x$ are points in the data space (observable representations of those cognitive states). Following the update rule $\mathcal{L}$, the latent space is traversed from $z_t$ to $z_{t+1}$, which is then projected to $x_{t+1}$ through the VAE decoder ($x_{t+1} = D_\mathcal{G}(z_{t+1})$), after which $x_{t+1}$ is provided to the classifier $\mathcal{D}$ to update the feedback.

Although the VAE is trained in an unsupervised way, having some labeled samples from the VAE's training data-set can be useful to understand latent space structure and track the progress of the simulated DecNef procedure. Each labeled example $\{x_i\in X^k\}_{i=1}^N$ from a category $k$ is mapped by the encoder to a Gaussian posterior distribution $q_\phi(z_i\mid x_i)$ with mean $z_i$ and covariance matrix $\sigma_i\mathbb{I}_{m}$, where $m$ is the dimension of the latent space and $\mathbb{I}_{m}$ is the identity matrix. From there, we define $z^k$ and $\sigma^k$ as the averages of the means and variances arising from the labeled samples from the category $k$, i.e. 

\begin{equation}\label{eq:prototype}
\text{``Latent prototype'' of class } k:= N(z^k, \sigma^k \mathbb{I}_{z_{dim}}) \text{ where}
    \begin{cases}
        z^k= \frac{1}{N}\sum\limits_{i=1}^N z_i \\
    \\
        \sigma^k= \frac{1}{N}\sum\limits_{i=1}^N \sigma_i 
    \end{cases}    
\end{equation}

$N(z^k, \sigma^k \mathbb{I}_{m})$ is called the latent prototype of class $k$, and its projection to the data space using the VAE's decoder (the ``prototype'' of class $k$) is representative of such category.

We have also evaluated the suitability of two incremental baseline models using a latent space with the same dimensionality $m$: the projection to the space of the $m$ first principal components using Principal Component Analysis (PCA), and a classical autoencoder (AE). PCA constitutes a linear feature transformation, and the AE adds nonlinear effects to the learned representations. Finally, the VAE contributes probabilistic sampling.

\subsubsection{Learning strategy}

In this example, we express the exploration/exploitation nature of the learning strategy by increasing/decreasing $\sigma$ in each iteration based on the current and previous DecNef feedback. The meaning of each model parameter is summarized in Table \ref{tab:parameters}. An increase in the reward signal favours a decrease in exploration (smaller $\sigma$), and viceversa. We model the changes in $\sigma$ as a linear combination of the current value and the reaction to the change in feedback, weighted by a parameter $\gamma$ representing the subject's reactivity or impulsivity, as parametrized by Equation \eqref{eq:sigma-update}
:

\begin{equation} \label{eq:sigma-update}
    \sigma_{t+1} = \left(1-\gamma \right) \sigma_t + \gamma \left(1-p_t\right)^2\left( \frac{p_{t-1}+\epsilon}{p_{t}+\epsilon}\right)^2\quad\text{for }t\geq1
\end{equation}

\noindent where $\epsilon = \max(p_t, p_{t-1})$ avoids division by zero if $p_t=0$.

To represent the assumption of stochastic exploration dynamics in self-regulation, we construct $z_{t+1}$  as a linear combination of the current state $z_t$ and a Gaussian term  $N(z_t, \sigma_{t+1})$, weighted by a parameter $\lambda$ which represents the subject's trust in the feedback. We encode the third assumption (reversal to the previous state $z_{t-1}$ under significantly worsening feedback) by modeling the learning strategy as a conditional function:

\begin{equation} \label{eq:update-rule}
    z_{t+1} = \mathcal{L}(z_t, p_t, p_{t-1}; \lambda, \gamma) =
    \begin{cases}
        (1-\lambda)z_t + \lambda N(z_t, \sigma_{t+1}) & \text{if } p_t > \delta p_{t-1}\\
        \\
        (1-\lambda)z_{t-1} + \lambda N(z_{t-1},\sigma_{t}) & \text{otherwise} 
    \end{cases}
    \quad\text{for }t\geq1
\end{equation}

\noindent where $\delta< 1$ marks the threshold where the loss of reward is such that the subject decides to revert to the previous state and attempt regulation again from there.

\begin{table}[t]
\centering
\begin{tabular}{l|l|p{7.5cm}}
\toprule
\textbf{Symbol} & \textbf{Type} & \textbf{Meaning} \\
\midrule
$z_t$ & Variable & Cognitive state of the subject at iteration $t$ (latent state).\\

$p_t$ & Variable & DecNef feedback (reward signal) obtained at iteration $t$.\\ 

$\sigma_t$ & Variable & Exploration scale at iteration $t$; larger values promote exploration, smaller values promote exploitation. \\

$\gamma$ & Parameter & Reactivity parameter controlling how strongly changes in feedback modulate the exploration scale $\sigma_t$. \\

$\lambda$ & Parameter & Trust-in-feedback parameter controlling the balance between the current state and stochastic exploration. \\

$\epsilon$ & Constant & Stabilization term preventing division by zero. \\

$\delta$ & Parameter & Feedback-rejection threshold ($\delta < 1$), determining when the subject reverts to the previous latent state due to significantly worsening feedback.\\\

$N(z, \sigma)$ & Distribution & Gaussian distribution centered at $z$ with standard deviation $\sigma$, modeling stochastic exploration. \\
\bottomrule
\end{tabular}
\caption{Summary of model parameters and variables present in Equations \ref{eq:sigma-update} and \ref{eq:update-rule}.}
\label{tab:parameters}
\end{table}

\section{Methods}

We conducted two parallel sets of DecNef simulations. In the first set (\textit{image experiments}), observable states were represented directly by Fashion-MNIST images, and both the classifier and generative model operated in image space. In the second set (\textit{synthetic fMRI experiments}), each Fashion-MNIST image was transformed into a corresponding synthetic fMRI activation pattern using MindSimulator~\cite{bao2025mindsimulator}. In this case, both the classifier and generative model operated exclusively on the resulting fMRI representations. Thus, the image and synthetic-fMRI experiments share the same underlying semantic categories but are trained and evaluated independently in different data modalities.

\subsection{Preparation and training} \label{sec:methods-training}

Using the FASHION-MNIST dataset~\cite{FASHION-MNIST}, which comprises 10 classes of $28 \times 28$ pixel images of apparel, we defined DecNef tasks in which the target state was represented by a specific image class. We considered two target classes: \emph{T-shirt/Top} (class 0) and \emph{Coat} (class 4). To investigate the impact of the alternative class used for classifier construction, we evaluated three alternative classes for each target. For the \emph{T-shirt/Top} target, the alternatives were \emph{Trouser} (class 1), \emph{Dress} (class 3), and \emph{Sneaker} (class 7). For the Coat target, the alternatives were \emph{Sandal} (class 5), \emph{Ankle Boot} (class 9), and \emph{Trouser} (class 1).

\subsubsection{Image experiments:}\label{sec:methods-preparation-images}

For the image experiments, we trained a binary classifier using images from the target and alternative classes. We used a Convolutional Neural Network (CNN)~\cite{lecun1989handwritten} with 4 convolutional blocks, each separated by a rectified linear unit (ReLu) activation function. The model was trained for 10 epochs using the Adam optimizer on the training split of FASHION-MNIST provided by the \verb|torchvision| library (version 0.16.0), restricted to the target and alternative classes.

In analogy to empirical DecNef, Fashion-MNIST samples from the target and alternative classes correspond to the participant's brain activity patterns evoked by different external stimuli during the classifier construction session.

As outlined in Section \ref{sec:model-definition}, the generative model was a VAE. It consisted of three convolutional blocks and a two-dimensional latent space, with a symmetric decoder architecture. The VAE was trained for 25 epochs using Adam~\cite{kingma2015adam} on images from all 10 classes present in the training split of FASHION-MNIST.

\subsubsection{Synthetic fMRI experiments:}\label{sec:methods-preparation-fmri}
Synthetic fMRI activation patterns were generated using MindSimulator~\cite{bao2025mindsimulator}, a pretrained image-to-fMRI model trained on image--fMRI pairs from the Natural Scenes Dataset (NSD)~\cite{allen2022massive}. Given an input image, MindSimulator predicts the voxel-wise brain activity that a specific NSD subject would be expected to exhibit when viewing that image, restricted to the visual cortex Region of Interest (ROI). We used the pretrained model corresponding to Subject 8 to transform all Fashion-MNIST images into synthetic fMRI activation patterns, which constituted the dataset used in all subsequent fMRI experiments. MindSimulator was employed exclusively as a preprocessing step without modification. The generated activations should therefore be interpreted as model-based approximations of subject-specific visual cortex responses. The validity of these representations relies on the evaluation reported by Bao et al.~\cite{bao2025mindsimulator}, who demonstrated substantial correspondence between predicted and experimentally measured NSD fMRI responses.

We then trained a binary ElasticNet logistic regression classifier with L1 and L2 regularization strengths $l_1=l_2=0.005$ using fMRI patterns corresponding to the target and alternative classes. Separately, we trained a VAE on fMRI patterns corresponding to all 10 Fashion-MNIST classes. To accommodate the higher dimensionality and complexity of fMRI data, the VAE employed a latent space of dimension $m=256$.

\subsection{Simulation of the induction stage} \label{sec:methods-induction}

The simulation of the DecNef induction stage began by selecting an initial latent state $z_0$. During $w=4$ warm-up iterations, the artificial participant was allowed to freely generate states without receiving meaningful feedback related to its current state. Since the learning strategy requires a probability value as input, we provided a constant feedback signal $\overset{static}{p_t}=0.5$ for $t=0,\ldots,3$.

Meanwhile, the latent states $z_t$ were decoded into observable samples $x_t$. Depending on the experiment, $x_t$ represented either Fashion-MNIST images or synthetic fMRI activation patterns. These samples were then evaluated by the corresponding pretrained classifier, yielding probabilities $p_t=p(y=y^\star \mid x_t)$.

The first meaningful feedback value shown to the participant was computed as the average classifier output obtained during the warm-up period, $\frac{1}{w}\sum_{t=0}^{w} p_t$. Subsequently, the participant reacted according to Equation \eqref{eq:update-rule} with $\lambda=\gamma=0.25$ and $\delta=0.75$. During the remainder of the simulation, feedback was supplied in real time using a moving average of the last $w$ classifier outputs. The process continued for $T=500$ iterations.

\subsection{Evaluation} \label{sec:methods-evaluation}

For evaluation purposes, we recorded the sequence of latent cognitive states induced during the simulation, $\{z_t\}_{t=0}^{T}$, constituting each cognitive trajectory. The decoder of the corresponding generative model produced the associated sequence of observable states $\{x_t\}_{t=0}^{T}$, where $x_t$ represented either Fashion-MNIST images or synthetic fMRI activation patterns depending on the experiment.

The image and synthetic-fMRI experiments are independent from each other, and were conducted analogously. In the image experiments, the decoder generated images that were evaluated by the image classifier. In the synthetic-fMRI experiments, the decoder generated synthetic fMRI activation patterns that were evaluated by the fMRI classifier. Therefore, the feedback signal always originated from a classifier operating in the same modality as the corresponding generative model. The classifier outputs $\{p_t\}_{t=0}^{T}$ constitute the neurofeedback signal and quantify the probability of the current state belonging to the target class.

As a control condition to verify whether DecNef learning was actually related to the feedback signal, we conducted equivalent simulations using feedback unrelated to either the current state or the target state. Specifically, we supplied $\overset{control}{p_t}\sim\text{Uniform}(0,1)$ independently of $x_t$.

In total, we conducted twelve experiments in each modality. For each of the two target classes, we evaluated three alternative classes and performed both a DecNef and a control simulation, yielding $2 \times 3 \times 2 = 12$ experiments per modality.

To facilitate interpretation of the synthetic-fMRI trajectories, we trained an auxiliary multilayer perceptron (MLP) on paired synthetic fMRI and image data generated with MindSimulator~\cite{bao2025mindsimulator}. The dataset consisted of pairs $(X,y)$, where $X$ denotes a synthetic fMRI activation pattern and $y$ denotes the corresponding Fashion-MNIST image used to generate it. The trained MLP maps voxel-wise fMRI representations back to image space, enabling visualization of the semantic content associated with generated fMRI states. This reconstruction model was used solely for post-hoc interpretation and visualization of trajectories and did not participate in the DecNef simulations, feedback computation, or quantitative evaluation.

\FloatBarrier

\subsubsection{Quality of latent representations compared to incremental baselines}\label{sec:methods-evaluation-representations}
\label{sec:baselines}

An important property of generative models used for DecNef simulation is that neighbouring points in the latent space should preserve their class identity. To analyze the incremental contribution of nonlinearity and probabilistic sampling to the quality of the learned latent representations, we compared the VAE against two alternative latent-space models with generative capabilities: a linear Principal Component Analysis (PCA) projection and a classical Autoencoder (AE). The comparison was performed separately for image and synthetic-fMRI data, which are independent sets of simulations.

All three generative models were trained on the same training data used in the corresponding DecNef experiments. Specifically, image-based models were trained and evaluated on Fashion-MNIST images, whereas fMRI-based models were trained on the synthetic fMRI activation patterns generated from those images. A multiclass classifier was also trained in each dataset.

To test latent space smoothness with respect to class identity for each generative model, we encoded the corresponding dataset to the latent representation and sampled 10000 points per class with replacement. We computed the standard deviation of the points from each class and added small perturbations using uniform noise with magnitude less or equal to 10\% of the standard deviation. We decoded the perturbed bootstrap samples and classified them using the multiclass classifier, computing AUC, Recall, Precision and F1. Supplementary Table \ref{tab:supp-baselines} shows that the VAE is superior in both datasets under study.

\FloatBarrier
\subsubsection{Analysis of the effect of initial conditions and stochastic effects} \label{sec:methods-randomness}

To assess the effect of the initial cognitive state on DecNef learning outcomes, we launched multiple simulations starting from different positions $z_0$ in latent space. For each modality, all experiments were run using the same set of initial states, thereby ensuring comparability within that modality. Specifically, we sampled 10 points from each Gaussian latent prototype distribution defined by Equation \eqref{eq:prototype}, resulting in 100 initial states. We additionally considered the stochastic effects introduced by the exploration dynamics of Equation \eqref{eq:update-rule}. To this end, we generated 10 trajectories from each initial state using different random seeds. Consequently, each experiment comprised 1000 trajectories in total. The same random seeds were used across all experiments to ensure trajectory-wise comparability.

\FloatBarrier
\subsubsection{Quantitative evaluation} \label{sec:methods-metrics}

We measured feedback maximization in terms of the percentage of potential improvement (or worsening) achieved for each DecNef trajectory, $\bar{L}$ (Equation~\eqref{eq:metric-L}), bounded in $[-100, 100]$. For a trajectory $i$ starting with feedback $p_{0i}$, the maximum possible improvement is $1-p_{i0}$ (reaching $p_{iT}=1$) and the maximum possible worsening is $p_0$. $\bar{L}_i$ measures the ratio of the actual improvement $p_{iT}-p_{i0}$ to the best/worst possible outcome distinguishing trajectories with final feedback greater/smaller than the initial feedback. We subtracted the same quantity obtained in the control experiments with random feedback to isolate the effect of DecNef training and divided by two to maintain the metric’s bounds between$ [-100, 100]$, defining $\Delta\bar{L}:=0.5\times(\bar{L}(\text{DecNef})-\bar{L}(\text{Random}))$.

In contrast with raw probability improvement, $\bar{L}$ takes into account the different maximum feedback increases possible for trajectories with different initial probability. This makes the metric comparable across individual trajectories within the same experiment. When averaged through all trajectories, $\bar{L}$ is also comparable across experiments and data modalities, which have experiment-specific distributions of initial feedback values due to each binary classifier producing a different feedback landscape.

\begin{equation}
    \bar{L}_i = 
    \begin{cases}
        100\times\frac{1}{T}\sum_{t=0}^T\frac{p_{it}-p_{i0}}{1-p_{i0}} & \text{if } p_{it} > p_{i0}\\[2ex]
        100\times\frac{1}{T}\sum_{t=0}^T\frac{p_{it}-p_{i0}}{p_{i0}} & \text{if } p_{it} \leq p_{i0} 
    \end{cases}
    \quad \text{with } i: \text{Trajectory index}
\end{equation}\label{eq:metric-L}

To inspect the properties of DecNef training in the cognitive space, we analyzed the relationship between feedback evolution and proximity to the target (both in the latent space and in the observable representation) using Pearson correlations for all trajectories. We also evaluated whether the decoder preserves target-related geometry by computing the Pearson correlation between distances to the target in latent space and distances to the target in observable space. We used euclidean distances in the latent space and pixel-by-pixel correlations in the observable space, which is applicable to both image and fMRI representations and measures similarity of activation patterns independently of global intensity differences.
\FloatBarrier

\section{Results}\label{sec:results}
\subsection{Results with image data}\label{sec:results-img}

In this section, we demonstrate how our simulation framework can be used to reflect on several questions commonly arising in DecNef research.  Questions \ref{sec:results-alternative-class-feedback} and \ref{sec:results-positive-feedback} can be answered by prior inspection of the generator and classifier, without the need to run any DecNef simulation. On the other hand, Questions \ref{sec:results-experimenters-perpective} and \ref{sec:results-z0-trajectory} are related to the dynamics of DecNef learning, and we explore them through systematic simulation of the DecNef loop under varying conditions.

Crucially, the artificial participant (defined in Section \ref{sec:model-definition}) is fixed throughout all the experimentation, which explores the variability in learning outcomes under different conditions, including experimental design choices and internal aspects.

\subsubsection{Does the choice of alternative class impact DecNef feedback?} \label{sec:results-alternative-class-feedback}

By definition, DecNef participants are unaware of the target during induction, and their brain patterns are driven solely by feedback. When, as usual, this feedback comes from a supervised classifier operating out of distribution, its target-class probabilities may be unreliable. Because supervised classifiers are comparative, the choice of alternative class affects the feedback. To examine this, we study how different different binary classifiers assign probabilities to the same target class across the VAE generator manifold.

We sampled $75\times75$ latent points $z=(z_x,z_y)$ from a square grid over the VAE latent space and decoded them to images $x$. Each image was then evaluated by binary classifiers for different target–alternative class pairs, yielding target-class probabilities $p(y=y^\star|x)$. The images were identical across evaluations; any variation in $p(y=y^\star|x)$ arose solely from the choice of alternative class.

\begin{figure}[h]
\centering
    \includegraphics[width=0.9\linewidth]{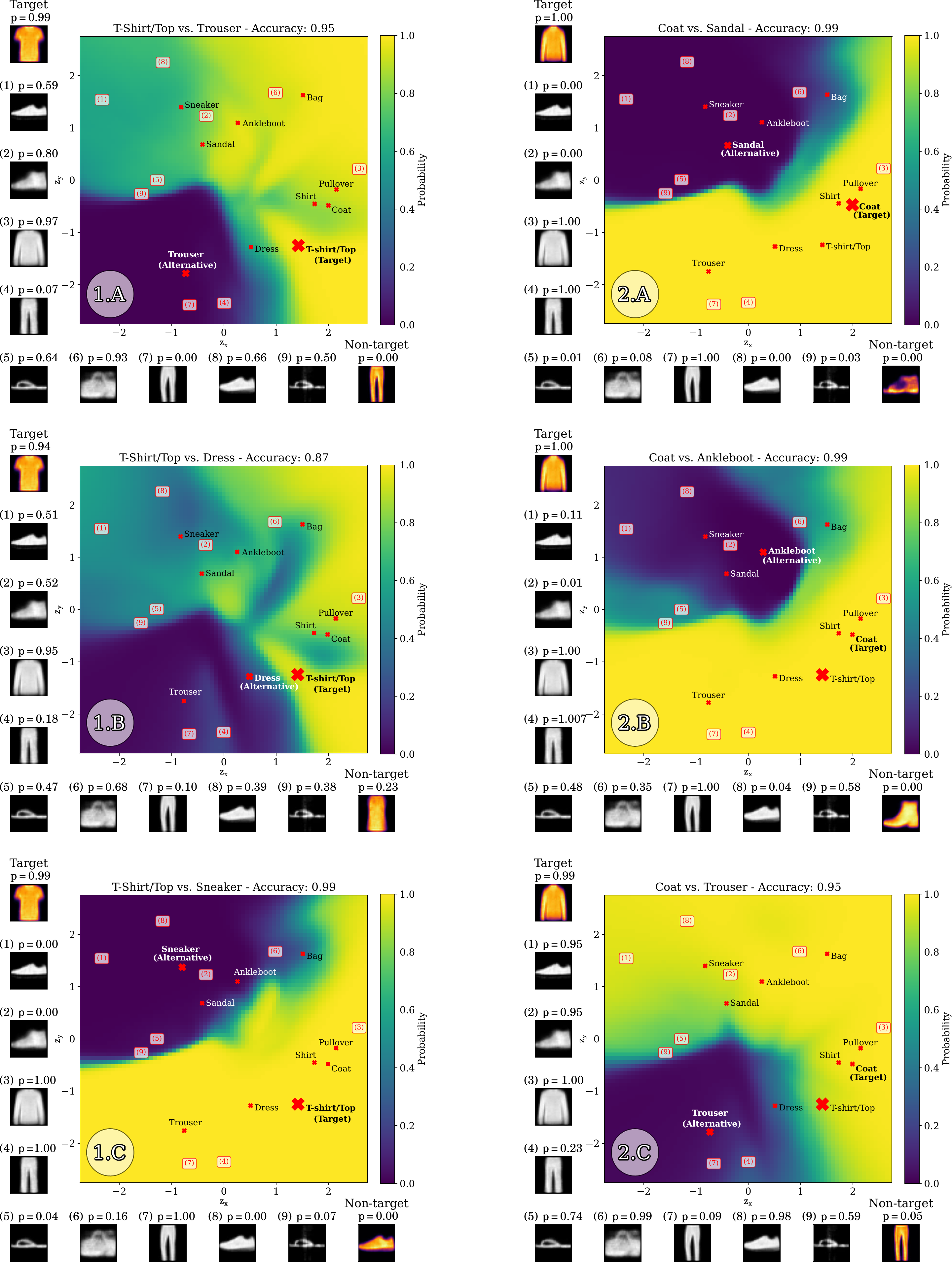}  
  \caption{Probabilities given by each binary discriminator to the images generated by the VAE via latent space sampling. Coordinates: $z_i$. Background Color: $p(y=y^\star \mid x_i)$, where $x_i$ is generated from $z_i$ using the VAE's decoder $D_\mathcal{G}$. Red markers: Location of the latent class prototypes. Warm-colored images: Class prototypes $x^\star$  and $x^{\mathrm{alt}}$ for the target and alternative classes. Grayscale images: Projections to the native space of the data of a random sample of latent coordinates, annotated by red numbers in the main panel and displaying $p(y=y^\star \mid x_i)$ above each grayscale image.}
\label{fig:pmaps}
\end{figure}

In each main panel of Figure \ref{fig:pmaps}, the background color shows the probability $p(y=y^\star \mid x)$ given by each binary classifier to the generated images. The red numbers annotated in the main panels mark a subset of randomly selected points $z_i$ in the latent space, and the smaller panels show the images $x_i$ generated by the decoder based on $z_i$, annotated with their corresponding $p(y=y^\star \mid x_i)$. The remarkable differences between the probability assigned to the same images by different binary classifiers applied out of their respective bimodal training distribution highlight the effect of the alternative-class choice, even with the same target-class.

\subsubsection{Does positive feedback imply successful target-class induction?}\label{sec:results-positive-feedback}

In Section \ref{sec:introduction}, we identify ``maladaptive learning'' as a key challenge in DecNef training. Prior work~\cite{laconte2011decoding, koush2012signal,heunis2020quality} suggests that participants may maximize feedback while substantially deviating from the target state, creating a false impression of success. Because a supervised classifier can be unreliable when applied to new contexts and data distributions, we adopt this view and further show that the choice of alternative class modulates this effect (Figure \ref{fig:pmaps}).

For instance, contrasting the target \emph{T-shirt/Top} against the alternative-class \emph{Sneaker} produces strongly positive feedback for most cognitive states, regardless of their true similarity to \emph{T-shirt/Top} (Panel 1C, predominantly yellow background), an effect also observed with \emph{Trouser} to a lesser degree (Panel 1A). In contrast, \emph{T-shirt/Top vs. Dress} restricts positive feedback to a small region (Panel 1B, predominantly blue/green background), and it assigns discouraging feedback to most states, raising further questions about discriminator design that will be explored in Section \ref{sec:results-z0-trajectory}. With \emph{Coat} as the target, all three alternative classes show strongly positive feedback for most classes out of distribution. Visualization of the feedback landscape anticipates maladaptative learning problems in all cases under study.

%\FloatBarrier
\subsubsection{Could initial conditions and stochastic effects bias experimenters toward labeling a participant as ``able'' or ``unable'' to learn DecNef?} \label{sec:results-experimenters-perpective}

As stated in Section \ref{sec:introduction}, a significant percentage of subjects (``non-responders'') do not benefit from DecNef. It remains worthwhile to investigate how momentary stochastic perturbations, accumulated across regulation attempts, interact with initial conditions to affect feedback maximization, implying that the same subject could succeed or fail in DecNef training depending on their initial state and broadening the perspective on ``non-responder'' characterization. Since real-life experiments cannot be repeated without interference from prior experience, we believe robust protocol design should account for this variability.

\begin{table}[htb]
    \centering
    \begin{tabular}{lcc}
            \toprule
             & $\bar{L}(\text{DecNef})$ &$\Delta\bar{L}(\text{DecNef})$\\
            \midrule
            T-shirt/Top vs Trouser & 28.05 & 0.83 \\
            T-shirt/Top vs Dress & 24.76 & 0.97 \\
            T-shirt/Top vs Sneaker & 39.14 & 8.56 \\
            Coat vs Sandal & 39.27 & 6.55 \\
            Coat vs Ankleboot & 40.77 & 5.34 \\
            Coat vs Trouser & 30.73 & 31.45 \\
            \bottomrule
        \end{tabular}
    \caption{Mean $\bar{L}$ per experiment, before and after subtracting the control condition. Units are percentages.}
    \label{tab:metric-L-img}
\end{table}

\begin{figure}[htbp]
    \centering
    \includegraphics[width=\linewidth]{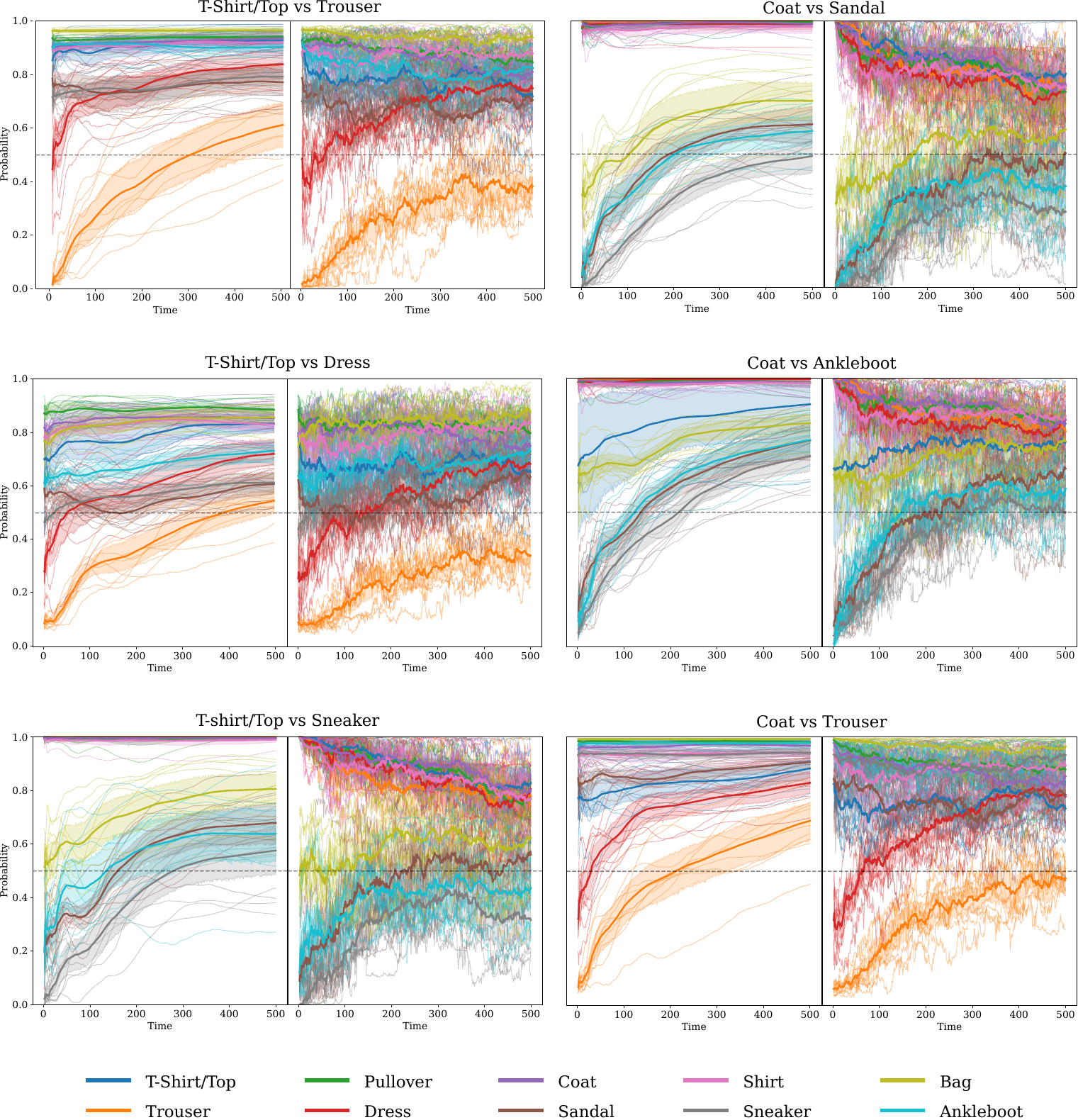}
\caption{Evolution of $p_t$ during the 1000 simulations for each experiment, grouped and averaged by initial point $z_0$ (thin lines). The color indicates the category of the Gaussian latent prototype from which each $z_0$ was sampled. The bold lines and shaded regions represent the average $p_t$ and 95\% confidence interval across all trajectories for which the initial point was sampled around the same latent prototype. Left column: Experiments with \emph{T-shirt/top} as the target. Right column: Experiments with \emph{Coat} as the target. Left panels for each experiment: DecNef simulation, with feedback given by the corresponding binary classifier. Right panels: Control condition with random feedback.} 
\label{fig:pt}
\end{figure}

As detailed in Section \ref{sec:methods-randomness}, we conducted a total of 1000 simulations per experiment to examine these issues. In Figure \ref{fig:pt}, we inspect the evolution of  $\{p_t\}_{t=0}^T$ in all trajectories, grouped by the initial state $z_0$ (100 thin lines) and color-coded by the starting region (bold lines and shaded regions: average $p_t$ and 95\% confidence interval per starting region), both in DecNef simulations (left panels for each experiment) and in the control simulations with random feedback (right panels).

We quantified performance in terms of maximum possible target probability increase/decrease to account for the effect of different initial feedback values, using $\bar{L}\in[-100, 100]$ --- Equation~\eqref{eq:metric-L}. Table \ref{tab:metric-L-img} shows the mean $\bar{L}$ across trajectories for the different experiments, before and after subtracting the control condition with random feedback. Controlling for the effect of random feedback reveals that learning is small to moderate. For the \emph{T-shirt/Top} target, alternative class \emph{Sneaker} produced the best results on average. For \emph{Coat} induction, the strongest alternative class, on average, was \emph{Trouser}, for which $\bar{L}(\text{Random})$ is negative. To test the null hypothesis that the alternative class used in DecNef does not affect feedback maximization performance, we conducted Kruskal–Wallis tests on the $\bar{L}$ values obtained for the three alternative classes corresponding to each target class. Statistical significance was obtained in all cases, with $p<0.001$.

\FloatBarrier
\subsubsection{How do these factors influence the evolution of internal cognitive states?} \label{sec:results-z0-trajectory} 

\begin{figure}[h]
\centering
        \includegraphics[width=0.8\linewidth]{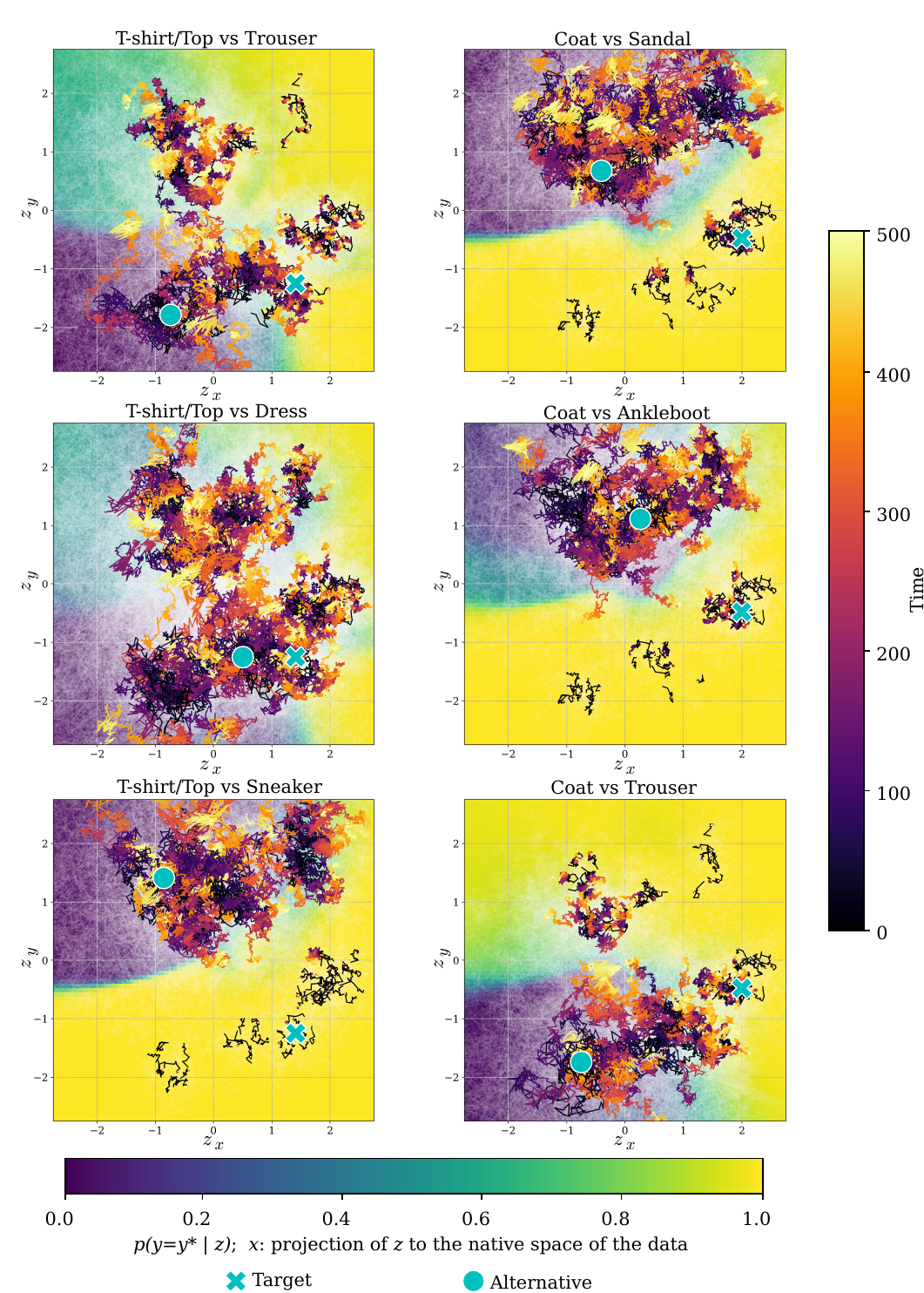}
\caption{Trajectories in the latent space for all DecNef simulations. Colored background: $p(y=y^\star \mid x)$ where $x=D_\mathcal{G}(z)$ for each coordinate $z$ and the VAE decoder denoted as $D_\mathcal{G}$. Thin white lines: all the individual trajectories. Bold lines: Average position of the 10 trajectories sharing the same $z_0$, with vertical color bar indicating time progression. Blue cross: Target class prototype. Blue circle: Alternative class prototype.}
\label{fig:trajs}
\end{figure}

In Section \ref{sec:introduction}, we have discussed how neuroimaging observations (i.e. fMRI scans) provide an indirect observation of the internal neural state, according to the \textit{decoder's dictum}, and how brain decoding alone does not guarantee knowledge about the content of the internal state. The DecNefSimulator framework underscores the need to clearly distinguish between the cognitive trajectory (in the latent space) and the observable trajectory (in the data space), highlighting that feedback depends on the observable trajectory, whereas the participants' attempts to regulate their own brain activity take place within the latent cognitive space.

The objective of DecNef is to teach the participant how to regulate $z_t$ by using $p_t$ to encourage a specific $Z^{\star}$. To that avail, $z_t$ is (indirectly) observed and its observable representation $x_t$ is obtained. Then, $x_t$ is interpreted through brain decoding methods (in DecNef, a supervised classifier), obtaining $p_t$, which is presented to the participant, prompting a change in $z_t$.

Our results in sections \ref{sec:results-alternative-class-feedback}-\ref{sec:results-experimenters-perpective} show that initial conditions, stochastic effects and classifier design affect successful reward maximization. Here, we inspect how those factors impact the development of the cognitive trajectory itself. Figure \ref{fig:trajs} shows the trajectories in the latent space for the different experiments, uncovering the internal cognitive states $z_t$ and also showing the associated performance measures $p_t$, derived from $x_t$. The colored background (bottom color bar) maps the probability given by the corresponding binary classifier to $x=D_\mathcal{G}(z)$ for each coordinate $z$ in the latent space. The thin white lines follow all the individual trajectories $\{z_t\}$, indicating the regions of the latent space that were explored during the simulation. The bold lines show the average position in time of each group of 10 trajectories sharing the same initial point $z_0$.

From Figure \ref{fig:trajs} we conclude that the initial cognitive state strongly influences the amount of exploration: trajectories departing from $z_0$ with a very high $p_0$ remain locally constrained, while trajectories with medium-to-low $p_0$ are more exploratory, and the feedback effectively prompts sequences of cognitive states that progressively increase the reward. This behaviour is observed consistently across experiments, with all panels displaying trajectories escaping low probability states (blue background) to finally reach high feedback regions (yellow background). However, the participant is only focused on reward maximization and displays no preference for regulation toward the actual desired state, and the trajectories that remarkably increase their feedback show no directional preference to approach the target (marked by a blue cross in each panel). Figure \ref{fig:trajs} also emphasizes the dependence on the alternative class. Changing the feedback landscape (colored background) changes affects the regions that are most likely to be explored or exploited, as well as the regions that, while being semantically different from the target, are encouraged by the feedback, offering a visualization of maladaptative learning.

Table \ref{tab:pearson-img} offers a quantitative analysis of these effects. Feedback is very weakly correlated with distance to the target state $z^\star$. In observable space, which is the actual domain of the classifiers, feedback is weakly/moderately correlated with similarity of $x$ to the observable representation $x^\star$ of the target, even though the decoder preserves target-related geometry (as interpreted from the positive correlation of distances $d(z, z^\star)$ and $d(x, x^\star)$). These findings suggest that classifier feedback does not reliably prompt movement toward the intended target state and may instead exploit alternative directions in representational space, increasing classifier confidence without achieving target induction.
\begin{table}[h]
    \centering
    \begin{tabular}{lrrr}
    \toprule
     & $p(y=y^\star | x)$ and $d(z, z^\star)$& $p(y=y^\star | x)$ and $\text{sim}(x, x^\star)$& $d(z, z^\star)$ and $d(x, x^\star)$\\
    \midrule
    T-shirt/Top vs Trouser & -0.03 & 0.14& 0.41 \\
    Coat vs Sandal & -0.00 & 0.19& 0.52 \\
    T-shirt/Top vs Dress & 0.12 & 0.16& 0.36 \\
    Coat vs Trouser & -0.01 & 0.04& 0.54 \\
    T-shirt/Top vs Sneaker & -0.09 & 0.27& 0.51 \\
    Coat vs Ankleboot & 0.03 & 0.20& 0.54 \\
    \bottomrule
    \end{tabular}
    \caption{Pearson correlations between feedback $p(y=y^\star | x)$ and distance to the target $d(z, z^\star)$, feedback and similarity to the observed representation of the target $\text{sim}(x, x^\star)$, and distances in the latent $d(z, z^\star)$ and the observable $d(x, x^\star)$ space. Distances in the latent space are euclidean, and similarity is the pixel-by-pixel correlation coefficient}
    \label{tab:pearson-img}
\end{table}

\FloatBarrier

\subsection{Results with synthetic fMRI}\label{sec:results-fmri}

\begin{figure}[h]
\centering
        \includegraphics[width=\linewidth]{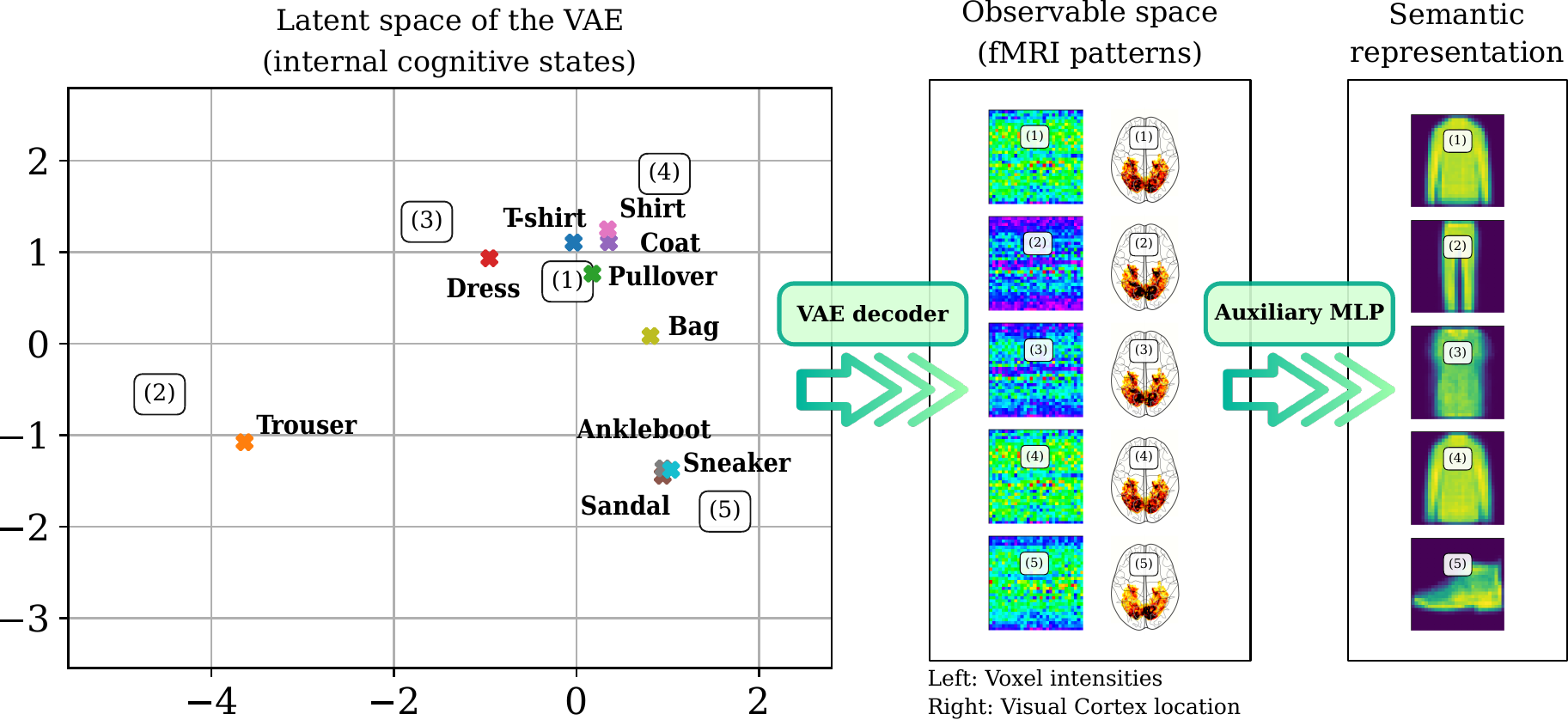}
\caption{Correspondence from the VAE latent space (in 2D, using PCA) to voxels in the brain and ultimately to clothing garments associated with those voxel responses. The latent prototypes are annotated in the first pannel. Five random points, (1)-(5), are ilustrated in their different representational spaces: As coordinates in the latent space (first panel), as voxel intensities in the observable space obtained by the VAE decoder (second panel), and as garments predicted by a MLP trained to generate images from fMRI patterns (third panel). The medium panel contains two visualizations of the same data: a square heatmap representation and a 3D representation that locates the voxels in the Visual Cortex.}
\label{fig:brain}
\end{figure}

In this section, we address the same questions posed in Section \ref{sec:results} in an additional set of simulations, now using synthetic fMRI data. The simulations are directly comparable to those conducted on image data, maintaining the experimentation pipeline intact. Since we used a high-dimensional VAE, dimensionality reduction is required for visualization. Moreover, the semantic content of the observed state in the fMRI space (i.e. the type of garment that elicited the fMRI pattern) is not directly accessible, but fMRI activations generated during the simulation can be visualized in brain maps and even translated to image space using a MLP, as detailed in Section \ref{sec:methods-evaluation}. Figure \ref{fig:brain} shows the location of the different class prototypes in the VAE latent space (left), signaling five specific cognitive states in PCA coordinates (points labeled A through E). Through the VAE decoder, the corresponding voxel activations for each of the five cognitive states and their location on the visual cortex are also displayed (middle). Finally, using the MLP, the semantic content of each cognitive state is uncovered (right).

\subsubsection{Does the choice of alternative class impact DecNef feedback?} \label{sec:results-alternative-class-feedback-fmri}

\begin{figure}[h]
\centering

    \includegraphics[width=\linewidth]{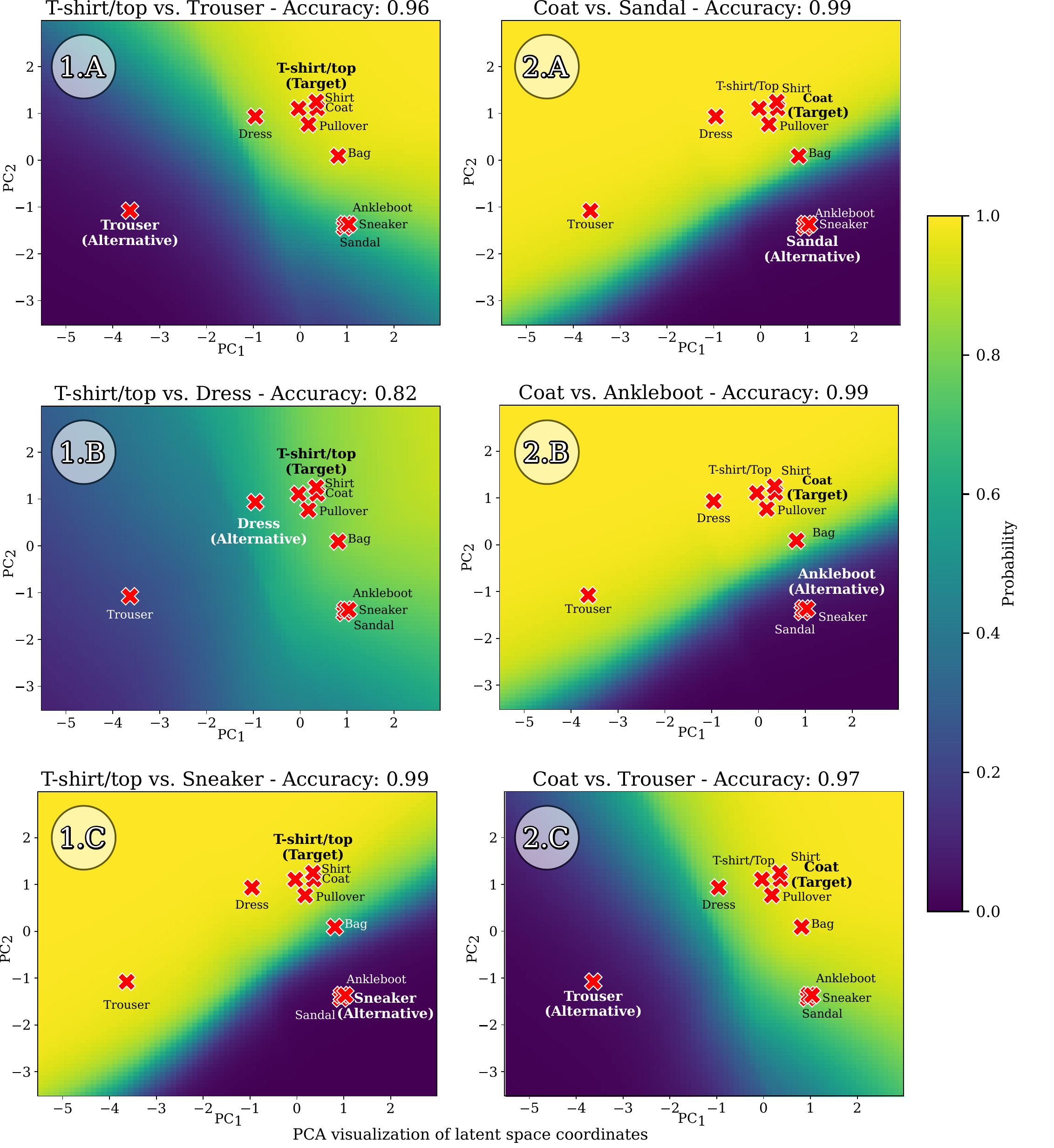}
 
  \caption{Probabilities given by each binary discriminator to the synthetic fMRI scans generated by the VAE via latent space sampling. Coordinates: First two Principal Components of $z_i$. Background Color: $p(y=y^\star \mid x_i)$, where $x_i$ is generated from $z_i$ using the VAE's decoder $D_\mathcal{G}$. Red markers: Location of the latent class prototypes. }
\label{fig:pmaps-fmri}
\end{figure}

To address this question, we conducted an identical analysis to the one presented in Section \ref{sec:results-alternative-class-feedback}. In this case, the latent space of the VAE was high-dimensional, so we used Principal Component Analysis (PCA) for visualization. To this end, PCA was fitted to the 10 latent class prototypes after standard normalization, and the data were projected onto the first two principal components. Then, we sampled points uniformly in a two-dimensional $75\times75$ grid and applied the inverse PCA transformation to obtain $\{z_i\}$, a set of points in the latent space of the VAE. These points were projected to the data space using the VAE decoder, yielding $\{x_i\}$, and the corresponding classifier was used to compute $p(y=y^\star \mid x_i)$, which is shown as the colored background in each panel of Figure  \ref{fig:pmaps-fmri}.  The annotations indicate the locations of the latent class prototypes in the space defined by the two leading principal components. The percentage of the variance explained by the first two principal components was 67.22\%.

\subsubsection{Does positive feedback imply successful target-class induction?}\label{sec:results-positive-feedback-fmri}

Similarly to Section \ref{sec:results-positive-feedback}, Figure \ref{fig:pmaps-fmri} reveals that positive feedback does not necessarily imply target class proximity, and that this issue depends heavily in the chosen classifier, with some classifiers (\emph{T-shirt/top vs Sneaker}, \emph{Coat vs Sandal}, \emph{Coat vs Ankleboot}) assigning high rewards to a wide range of non-target states and anticipating maladaptive learning.

\FloatBarrier
\subsubsection{Could initial conditions and stochastic effects bias experimenters toward labeling a participant as ``able'' or ``unable'' to learn DecNef?} \label{sec:results-experimenters-perpective-fmri}

\begin{figure}[h!]
    \centering
    \includegraphics[width=\linewidth]{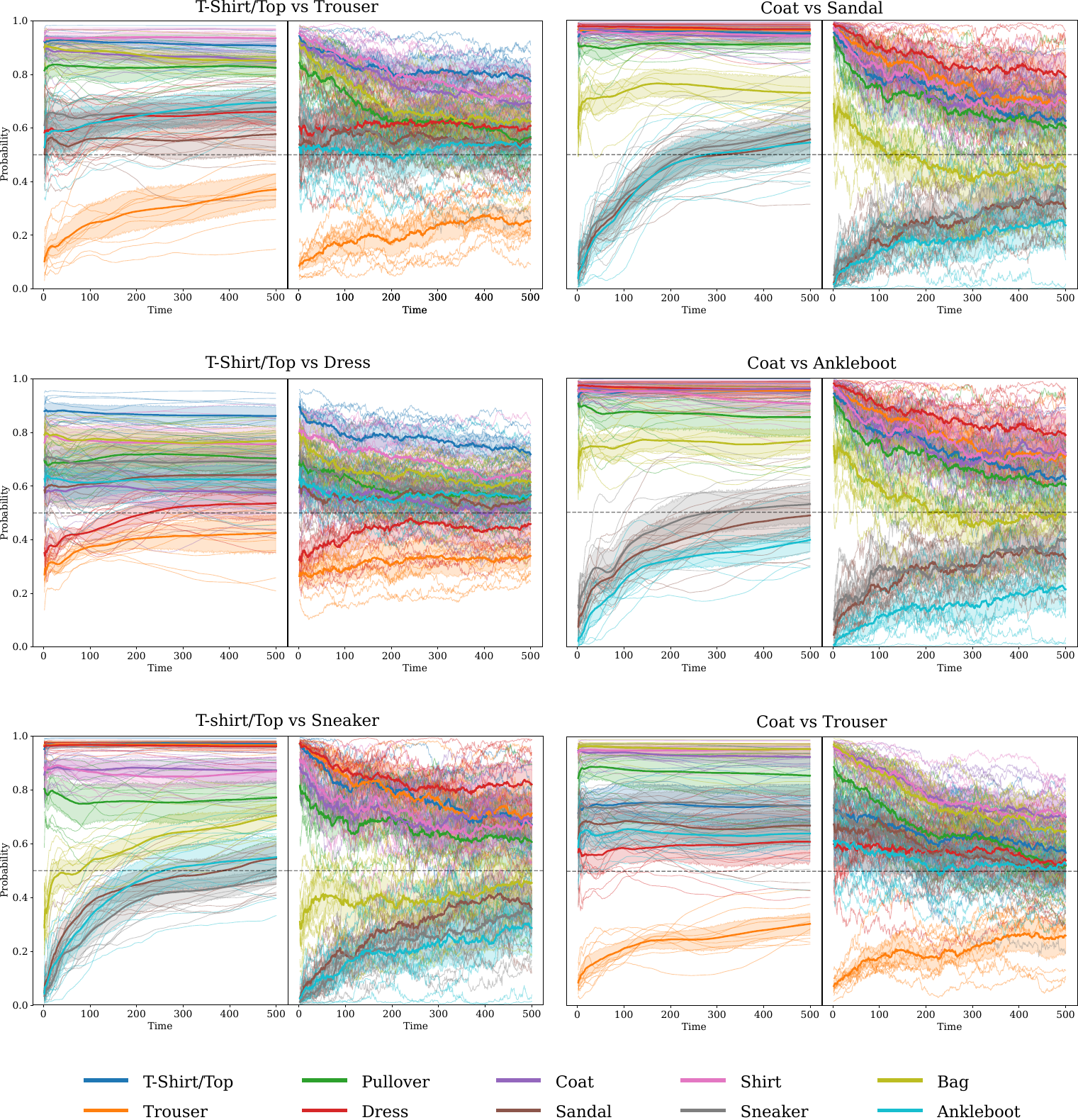}
\caption{Evolution of $p_t$ during the 1000 simulations for each experiment, grouped and averaged by initial point $z_0$ (thin lines). The color indicates the category of the Gaussian latent prototype from which each $z_0$ was sampled. The bold lines and shaded regions represent the average $p_t$ and 95\% confidence interval across all trajectories for which the initial point was sampled around the same latent prototype. Left column: Experiments with \emph{T-shirt/top} as the target. Right column: Experiments with \emph{Coat} as the target. Left panels for each experiment: DecNef simulation, with feedback given by the corresponding binary classifier. Right panels: Control condition with random feedback.} %
\label{fig:pt-fmri}
\end{figure}

In Figure \ref{fig:pt}, we inspect the evolution of  $\{p_t\}_{t=0}^T$ in all trajectories, grouped by the initial state $z_0$ (100 thin lines) and color-coded by the starting region (bold lines and shaded regions: average $p_t$ and 95\% confidence interval per starting region), both in DecNef simulations (left panels for each experiment) and in the control simulations with random feedback (right panels). For this analysis, we followed the procedure detailed in Section \ref{sec:methods-randomness}, identical to the experiments with images (presented in Section \ref{sec:results-experimenters-perpective}). Therefore, Figure \ref{fig:pt-fmri} shows the evolution of  $\{p_t\}_{t=0}^T$ in all trajectories, grouped by the initial state and color-coded by the starting region, in DecNef (left panels for each experiment) and control conditions (right panels) .

Performance was measured in terms of maximum possible target probability increase/decrease,  $\bar{L}\in[-100, 100]$ (Equation~\eqref{eq:metric-L}). Table \ref{tab:metric-L-fmri} shows the mean $\bar{L}$ across trajectories for the different experiments, before and after subtracting the value obtained in control simulations. Learning is small to moderate. For the \emph{T-shirt/Top} target, alternative class \emph{Sneaker} produced the best results on average, in accordance with the image simulation. For \emph{Coat} induction the strongest alternative class was \emph{Sandal}, on average, which differs from the result obtained with images. We applied Kruskal–Wallis tests to the $\bar{L}$ values for the three alternative classes associated with each target class, testing the null hypothesis that the choice of alternative class in DecNef does not influence feedback maximization performance. In every case, the results were statistically significant, with $p<0.001$.

\begin{table}[h]
    \centering
        \begin{tabular}{lrr}
        \toprule
         & $\bar{L}(\text{DecNef})$ & $\Delta\bar{L}(\text{DecNef})$\\
        \midrule
        T-shirt/Top vs Trouser & 17.53 & 8.33 \\
        T-shirt/Top vs Dress & 11.22 & 5.82 \\
        T-shirt/Top vs Sneaker & 30.10 & 11.18 \\
        Coat vs Sandal & 24.36 & 14.37 \\
        Coat vs Ankleboot & 20.06 & 8.62 \\
        Coat vs Trouser & 17.48 & 7.22 \\
        \bottomrule
        \end{tabular}
    \caption{Mean $\bar{L}$ per experiment, before and after subtracting the control condition.}
    \label{tab:metric-L-fmri}
\end{table}

\FloatBarrier
\subsubsection{How do these factors influence the evolution of internal cognitive states?} \label{sec:results-z0-trajectory-fmri} 
\FloatBarrier
\begin{figure}[h]
\centering
        \includegraphics[width=\linewidth]{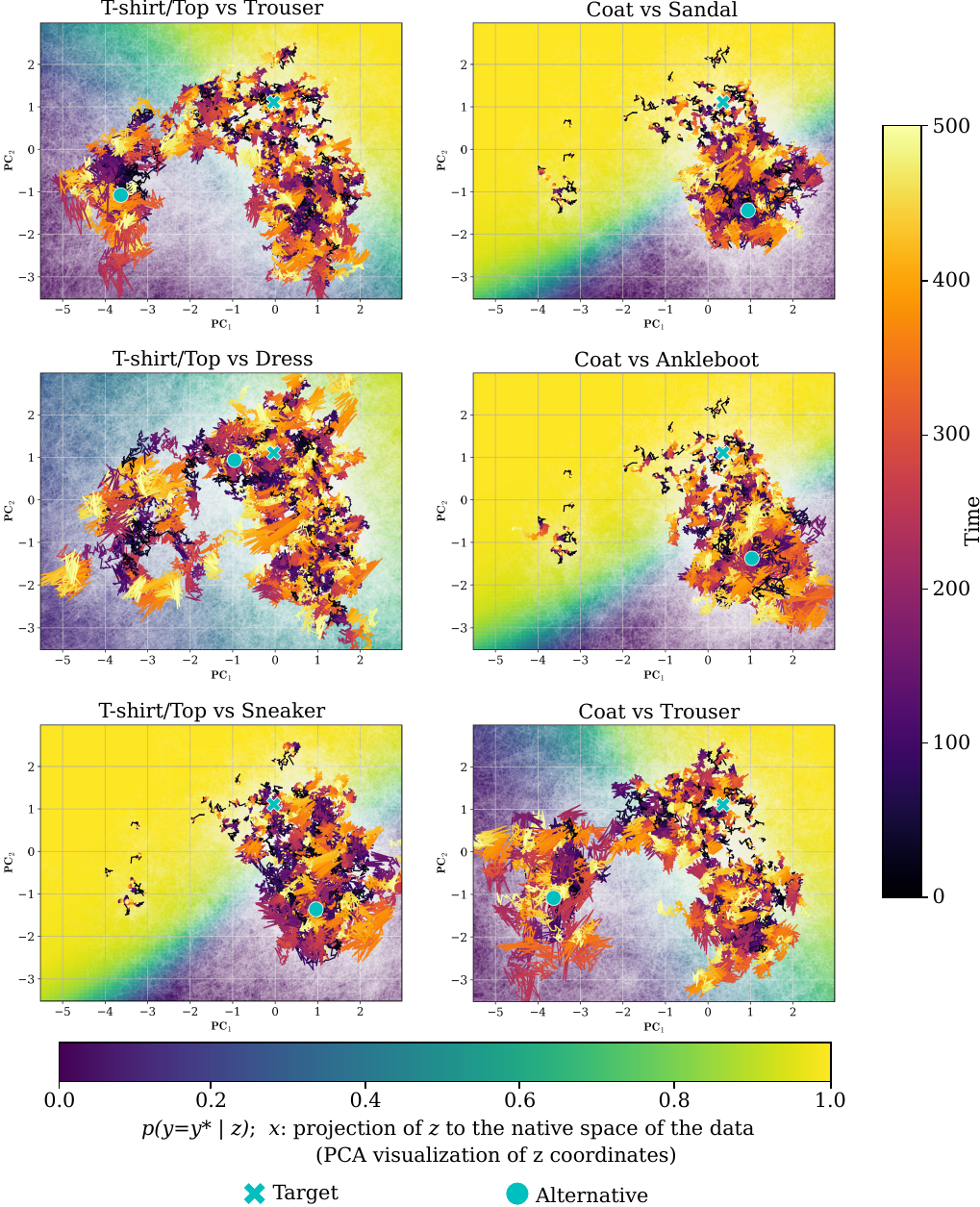}
\caption{Trajectories in the latent space for all DecNef simulations. Colored background: $p(y=y^\star \mid x)$ where $x=D_\mathcal{G}(z)$ for each coordinate $z$ and the VAE decoder denoted as $D_\mathcal{G}$. Thin white lines: all the individual trajectories. Bold lines: Average position of the 10 trajectories sharing the same $z_0$, with vertical color bar indicating time progression. Blue cross: Target class prototype. Blue circle: Alternative class prototype.}
\label{fig:trajs-fmri}
\end{figure}

As mentioned in Section \ref{sec:results-z0-trajectory}, DecNefSimulator allows tracking cognitive trajectories transparently instead of relying merely in the observed trajectory, bypassing the ``decoder's dictum''.  When using high-dimensional cognitive spaces, additional dimensionality reduction methods are required for visualization, but the development of cognitive trajectories can still be observed.

Figure \ref{fig:trajs-fmri} shows the projection of the cognitive trajectories onto the 2D PCA space, mapping the internal cognitive states $z_t$ to their associated performance measures $p_t$, derived from $x_t$ (colored background). It can be observed that exploration is less likely when the initial $p_0$ is high. In cognitive regions mapping to observable states with poor reward signal, the feedback successfully modifies the trajectory to induce cognitive states mapping to observations with higher $p(y=y^{\star} \mid x)$, but no preference is shown towards actually approaching the desired state.

\begin{table}[h]
    \centering
    \begin{tabular}{lrrr}
    \toprule
     & $p(y=y^\star | x)$ and $d(z, z^\star)$& $p(y=y^\star | x)$ and $\text{sim}(x, x^\star)$& $d(z, z^\star)$ and $d(x, x^\star)$\\
    \midrule
    T-shirt/Top vs Trouser & 0.26 & -0.02 & 0.46 \\
    T-shirt/Top vs Dress & 0.14 & 0.03 & 0.53 \\
    T-shirt/Top vs Sneaker & 0.45 & 0.05 & 0.33 \\
    Coat vs Ankleboot & 0.43 & -0.01 & 0.41 \\
    Coat vs Trouser & 0.28 & 0.07 & 0.44 \\
    Coat vs Sandal & 0.49 & 0.01 & 0.35 \\
    \bottomrule
    \end{tabular}
    \caption{Pearson correlations between feedback $p(y=y^\star | x)$ and distance to the target $d(z, z^\star)$, feedback and similarity to the observed representation of the target $\text{sim}(x, x^\star)$, and distances in the latent $d(z, z^\star)$ and the observable $d(x, x^\star)$ space. Distances in the latent space are euclidean, and similarity is the pixel-by-pixel correlation coefficient}
    \label{tab:pearson-fmri}
\end{table}

Table \ref{tab:pearson-fmri} quantifies these effects. In all the experiments, feedback is consistently associated with increased distance from the target latent prototype, despite the fact that latent and observable distances remain positively coupled. Positive correlation between feedback and $\text{sim}(x, x^\star)$ indicates that the decoder preserves target-related geometry. These findings indicate that classifiers are exploiting alternative directions to increase confidence without moving toward the intended target state, behaving in an adversarial manner. Compared to the image experiments, this problem is accentuated even further due to the high-dimensional latent space.

\FloatBarrier
\section{Discussion}\label{sec:discussion}
Decoded Neurofeedback (DecNef) is a non-invasive technique for brain modulation with a growing range of applications in both clinical interventions and cognitive enhancement~\cite{koizumi2016fear, cortese2017decoded}. It is also a valuable research tool to increase our understanding about brain function ~\cite{amano2016learning, shibata2021mechanisms, cortese2020unconscious, margolles2024unconscious}. However, DecNef is nowadays still hindered by concerning limitations~\cite{yamashita2008sparse, cortese2016multivoxel, heunis2020quality, shibata2021mechanisms}, namely, undesirable reinforcement of cognitive states deviating from the target, domain shift between the discriminator training stage and the target neural pattern of DecNef during induction~\cite{yamashita2008sparse,cortese2016multivoxel,olza2025domain} or the subject-dependency of learning~\cite{alkoby2018can, haugg2021predictors}. These limitations encourage the development of a comprehensive simulation framework to understand and improve DecNef in a controlled environment. In this work, we introduce a framework designed for this purpose and demonstrate its utility through a series of computational simulations.

Our simulations revealed three main findings: (i) the choice of alternative class used by the classifier during feedback computation is a critical determinant of learning outcomes; (ii) the perceived (in)ability to learn can be driven by circumstantial factors, such as cumulative randomness in transient decisions, which may lead experimenters to label a participant as a ``non-responder''; and (iii) the initial cognitive state strongly shapes the sequence of states elicited during induction.

Beyond their theoretical implications, these findings have practical consequences for the design of real DecNef protocols. First, they suggest that the stimuli used during decoder construction should not be selected solely based on experimental convenience, since the choice of alternative class can substantially alter the feedback landscape and subsequent learning outcomes. Particularly promising is the replacement of supervised discriminators with self-supervised or unsupervised approaches that directly model the target class, avoiding issues introduced by comparative feedback computation. Second, the strong dependence on the initial cognitive state indicates that participant-specific calibration procedures may improve learning consistency by guiding participants toward favorable regions of the neural state space before induction begins. Third, the proposed simulation framework provides a means to evaluate and optimize protocol design choices in silico before conducting costly neuroimaging experiments, potentially reducing unsuccessful experimental iterations and facilitating the development of more robust neurofeedback interventions.

Previous simulation studies~\cite{oblak2017selfregulation, shibata2019toward, annicchiarico2025activeinferenceperspectiveneurofeedback} have focused on specific aspects of DecNef simulation, such as demonstrating neural plasticity~\cite{shibata2019toward} or modeling the influence of prior beliefs and confidence~\cite{annicchiarico2025activeinferenceperspectiveneurofeedback}, but have limited flexibility and generalizability. Moreover, they have not explored interactions among cognitive states associated with semantically related concepts. In contrast, our work delves directly into those aspects, emphasizing modularity and task-flexibility  and unlocking direct access to the internal dynamics of the induction process.

DecNefSimulator explicitly considers the methodological constraints of neuroimaging-based decoding. By computing the classifier's output $p(y=y^{\star} \mid x)$ from observable proxies $x$ (analogous to fMRI data), we can test the validity of the \textit{decoder's dictum}: the assumption that successful external decoding reflects the functional use of information by the brain~\cite{deWit2016neuroimaging}. Unlike empirical DecNef experiments, where the true cognitive state remains hidden, DecNefSimulator decouples the experimenter's perspective from the latent dynamics $z$, allowing direct assessment of whether maximizing $p(y=y^{\star} \mid x)$ genuinely drives regulation toward the desired cognitive state $z^{\star}$. Consequently, the framework enables analysis of causal, rather than merely correlational, relationships between feedback and cognitive state changes.

Once the latent representation has been learned, the simulation proceeds entirely within latent space and no longer requires additional external data. Consequently, the framework can be used for other neuroimaging modalities, such as EEG or MEG, provided that suitable generative models and feedback mechanisms are available for the characteristics of the data. In our experimental demonstration, image-based simulations facilitate semantic interpretation, whereas synthetic fMRI simulations provide greater realism for DecNef applications.

Our computational experiments illustrate several key capabilities of the framework. The update rule $\mathcal{L}$, encoding participant behavior, can be easily adapted to represent different cognitive strategies. For example, we modeled a ``cautious'' participant who reverts to previous states following negative feedback, but alternative formulations could capture more exploratory or directional behavior. The parameters $\lambda$ and $\gamma$ respectively control confidence and attention to the feedback signal, allowing systematic investigation of their influence on learning dynamics. As shown in our binary classifier comparisons, the framework also exposes potential design flaws; for instance, how poor alternative class choices can impede learning by shaping the reward landscape.

These simulations highlight subtle yet critical aspects of DecNef learning. Trajectories starting in high-reward regions remain locally constrained, whereas trajectories beginning in low-reward regions often maximize feedback without converging toward the target state. Moreover, changing the alternative class dramatically alters the feedback topology and learning dynamics, underscoring how classifier construction critically shapes participant behavior. Such findings suggest that some subjects labeled as ``non-learners'' may, in fact, be constrained by suboptimal experimental design and alternative class choice rather than inherent inability.

Importantly, the present version of DecNefSimulator is not intended to provide  a neurobiologically detailed account of cognition or to capture the full richness of an individual's cognitive processes. Instead, its purpose is to provide a generative framework for investigating neural induction of semantic categories. Consequently, the participant-level latent variable model should be interpreted as a task-specific representation that encodes semantic structure relevant to the experimental objective, rather than as a generalized, exhaustive model of an individual’s cognitive architecture across multiple domains of brain function. [...] Future research should explore whether our work can be applied to a broader range of cognitive and affective processes, as well as to more ecologically valid settings, where additional constraints and more sophisticated models of participant behavior may be required. Given the modular nature of our framework, integration with state-of-the-art models of the brain~\cite{serin2025generating, dascoli2026foundation} arises as a natural direction for future development.

The simplicity of the learning strategy presented offers a transparent and interpretable account of DecNef learning, combining three assumptions grounded in cognitive neuroscience. The stochastic transitions modeled through $\mathcal{N}(z_t,\sigma_{t+1})$ are consistent with noisy neural population dynamics~\cite{ueltzhoffer2015stochastic} and with theoretical accounts of Gaussian optimality for exploration-exploitation balance in continuous RL~\cite{wang2019exploration}, with exploration reflected in the mean and exploitation in the variance of the optimal Gaussian distribution. Feedback modulation of exploration through Equation~\eqref{eq:sigma-update} mirrors the adaptive exploration behaviour documented in ~\cite{phillips2011implied}. The quadratic amplification of the ratio $p_{t-1}/p_t$ in \eqref{eq:sigma-update} and the modulation term $(1-p_t)^2$ are also consistent with the Pearce-Hall model of uncertainty-driven learning as studied in~\cite{queirazza2019neural}, and with the adaptive noise dynamics of the Ornstein-Uhlenbeck Adaptation framework~\cite{garciafernandez2024ornstein}. Within this interpretation, $\gamma$ controls the rate at which exploration adapts to feedback, a characteristic whose neural correlate is the precision of dopaminergic signalling in the striatum~\cite{friston2014dopamine}. $\lambda$ models the participant's trust in the feedback signal, analogous to precision weighting in Active Inference~\cite{friston2016active}. Reverting to the previous state after a large reward decrease implements a simple credit-assignment mechanism analogous to the ``stay-or-switch'' logic proposed in neurofeedback importance-sampling models~\cite{davelaar2018mechanisms}. Although simplified, this formulation captures core aspects of exploratory learning while remaining easy to interpret and modify.

Several extensions of the framework are possible. Alternative formulations of the learning rule could incorporate memory of past cognitive states, directional exploration, softmax Q-learning~\cite{sutton2018reinforcement}, actor-critic approaches~\cite{makoto2011multiple}, or Active Inference models~\cite{friston2016active}. While these alternatives carry greater theoretical generality, they also introduce substantially higher complexity and reduced interpretability, motivating the simpler and more interpretable closed-form rule adopted here. Owing to the modular design of the framework, such enhancements can be introduced without altering its core structure.

Likewise, the generator $\mathcal{G}$ could be replaced by alternative latent-variable architectures, including Adversarially Learned Inference models~\cite{dumoulin2017adversarially}, Hierarchical VAEs~\cite{havtorn2021hierarchical}, Variational Diffusion Models~\cite{kingma2021variational}, recent foundation models of brain activity~\cite{dascoli2026foundation}, or fMRI generators tailored to synthetize task-based contrast maps~\cite{serin2025generating}. Importantly, $\mathcal{G}$ is not intended as a perfect reconstructor but as an imperfect model of internal representations, analogous to how mental imagery provides a noisy reflection of perception. While previous work suggests that latent generative models capture meaningful structure in brain activity~\cite{takagi2023high,kamitani2025visual}, a systematic evaluation of alternative latent representations remains an important direction for future research. The recent foundational model for brain activity prediction across stimuli modalities presented in~\cite{dascoli2026foundation} could potentially be adapted for integration within our framework, although the properties of the model's latent space have to be carefully assessed. The work in~\cite{dascoli2026foundation} does not constitute a simulation framework for DecNef, since it does not explore transitions between cognitive states, but analyzing the semantic coherence of the model's internal layers (and, importantly, the ability to interpolate between concepts) is an interesting direction for future work.

In summary, DecNefSimulator introduces a flexible, transparent, and task-flexible framework for DecNef simulation that bridges cognitive-level modeling of semantic category induction and empirical neurofeedback research. By revealing the full trajectory of learning and allowing causal inspection of the relationship between the feedback and the cognitive state, it provides an unprecedented opportunity to refine DecNef methodology. We envision DecNefSimulator as both a tool for methodological innovation and a conceptual bridge toward more reproducible, interpretable, and causally grounded DecNef research.

\section{Conclusions}

We have introduced an adaptable and task-flexible simulation framework for DecNef that allows experimenters to explore and refine neurofeedback protocols \textit{in silico} before applying them to human participants. Unlike existing biologically or cognitively focused approaches, our method provides direct access to the induced states and the learning dynamics that drive them, making the DecNef process fully transparent and analyzable.

By offering modular components for generators, discriminators, and update rules, the framework accommodates a wide range of experimental designs and supports systematic evaluation of protocol parameters and potential design. This enables researchers to identify both effective strategies and potential failures, reducing the trial-and-error burden of \textit{in vivo} experimentation.

Ultimately, this work contributes a new tool for the neurofeedback community: one that bridges the gap between computational modeling and experimental practice, facilitates methodological innovation, and lays the groundwork for more robust and efficient development of DecNef protocols.

\section{Acknowledgements}
Supported by the University of the Basque Country via Grant PIF2025 UPV/EHU, by the Spanish Ministry of Science and Innovation (Projects PID2023-149195NB-I00 and PID2022-137442NB-I00), by Elkartek 
(KK-2023/00012, KK-2023/00090, KK-2024/00030), by the Spanish Ministry of Economy and Competitiveness, through the ``Severo Ochoa'' Programme for Centres/Units of Excellence (CEX2020-001010-S) and also from project grant 
PID2019-105494GB-I00. DS acknowledges funding by the Spanish State Research Agency through BCBL Severo Ochoa excellence accreditation CEX2020-001010/AEI/10.13039/501100011033 and through project PID2023-149267NB-I00.

\FloatBarrier

%% Loading bibliography style file
%\bibliographystyle{model1-num-names}
\bibliographystyle{elsarticle-num}

% Loading bibliography database
%\bibliography{cas-refs}
\bibliography{full_bibliography}
% Biography
%\bio{}
% Here goes the biography details.
%\endbio

%\bio{pic1}
% Here goes the biography details.
%\endbio
\FloatBarrier
\appendix

\section{Supplementary analyses}\label{app:experiments}
\FloatBarrier
\begin{table}[h]
    \centering
    \begin{tabular}{c|cc|cc|cc|cc}
    & \multicolumn{2}{c}{AUC (OvR)} &  \multicolumn{2}{c}{Recall} &  \multicolumn{2}{c}{Precision} &  \multicolumn{2}{c}{F1} \\
    %\cmidrule{2-9}
    \midrule
   & Images & fMRI & Images & fMRI & Images & fMRI & Images & fMRI \\ 
PCA & 0.8260 & 0.9207 & 0.3177 & 0.5194 & 0.2162 & 0.5968 & 0.2457 & 0.4875 \\
AE  & 0.9394 & 0.9209 & 0.5806 & 0.5057 & 0.6316 & 0.5965 & 0.5371 & 0.4702 \\
VAE & \textbf{0.9417} & \textbf{0.9213} & \textbf{0.5991} & \textbf{0.5248} & \textbf{0.7395} & 0.5970 & \textbf{0.5578} & \textbf{0.4927} \\
p-value (ANOVA) & $< 10^{-4}$ & 0.0118 & $< 10^{-4}$ & $< 10^{-4}$ & $< 10^{-4}$ & 0.8899 & $< 10^{-4}$ & $< 10^{-4}$\\
    \end{tabular}
    \caption{Performance of a multiclass classifier on perturbed points sampled from the latent space of three generative models (PCA, AE and VAE).}
    \label{tab:supp-baselines}
\end{table}
\FloatBarrier

\clearpage
\section{Computational requirements}\label{app:computational}
\begin{table}[h]
    \centering
    \resizebox{\textwidth}{!}{
    \begin{tabular}{c|c|c|c|c|c|c|c}
    Feedback & Dataset & Target & Alternative  & VAE training & Classifier training &  Trajectory (mean) & Total\\
     \midrule
      \multirow{8}{*}{DecNef} & \multirow{4}{*}{Images} & T-shirt/top & Trouser & \multirow{4}{*}{361s} & 44s & 7.59s & 2h 6min 31s \\
       & & T-shirt/top & Dress &  & 42s & 7.78s & 2h 9min 39s \\
       & & T-shirt/top &Sneaker&  & 44s & 7.35s & 2h 2min 39s \\
       & & Coat   & Ankle boot &  & 40s & 7.49s & 1h 52min 23s \\
       & & Coat   & Sandal     &  & 43s & 7.42s & 2h 3min 45s \\
       & & Coat   & Trouser    &  & 41s & 7.41s & 2h 3min 33s \\
      \cmidrule{2-8}
      & \multirow{4}{*}{fMRI} & T-shirt/top & Trouser & \multirow{4}{*}{130s} & 9s & 7.21s & 2h 6min 31s \\
      & & T-shirt/top & Dress &  & 8s & 6.97s & 1h 56min 15s \\
      & & Coat   & Ankle boot &  & 7s & 6.93s & 1h 55min 41s \\
      & & Coat   & Sandal     &  & 8s & 6.94s & 1h 55min 50s \\
      & & Coat   & Trouser    &  & 9s & 7.32s & 2h 2min 3s  \\
      \midrule
      \multirow{8}{*}{Random} & \multirow{4}{*}{Images} & T-shirt/top & Trouser & \multirow{4}{*}{} &  & 5.73s & 1h 35min 32s \\
       & & T-shirt/top & Dress &  &  & 5.72s & 1h 35min 21s \\
       & & T-shirt/top &Sneaker&  &   & 6.93s & 1h 55min 32s \\ 
       & & Coat   & Ankle boot &  &  & 5.79s & 1h 26min 51s \\
       & & Coat   & Sandal     &  &  & 5.74s & 1h 35min 40s \\
       & & Coat   & Trouser    &  &  & 5.74s & 1h 35min 39s \\
      \cmidrule{2-8}
      & \multirow{4}{*}{fMRI} & T-shirt/top & Trouser & \multirow{4}{*}{} &  & 6.96s & 1h 56min 4s \\
      & & T-shirt/top & Dress &  &  & 6.98s & 1h 56min 24s \\
      & & T-shirt/top &Sneaker&  &   & 6.78s & 1h 53min 4s  \\ 
      & & Coat   & Ankle boot &  &  & 6.86s & 1h 54min 16s \\
      & & Coat   & Sandal     &  &  & 6.89s & 1h 54min 46s \\
      & & Coat   & Trouser    &  &  & 6.88s & 1h 54min 37s \\
    \end{tabular}
    }
    \caption{Runtimes on a NVIDIA GeForce RTX 2080 GPU. The full experiment consists of classifier training, VAE training and 1000 trajectories with 500 iterations each. Each VAE and classifier are trained only once. The amount of trajectories and iterations can be changed as necessary for faster experimentation.}
    \label{tab:placeholder}
\end{table}
\FloatBarrier

\end{document}